# Mono-elemental saturable absorber in mode-locked fiber laser: A review

Kuen Yao Lau[1,*], Jian-Cheng Zheng[2], Cuihong Jin[1], and Song Yang[3]

[1] *State Key Laboratory of Modern Optical Instrumentation, College of Optical Science and Engineering, Zhejiang University, Hangzhou 310027, China.*

[2] *College of Electronic and Information Engineering, Nanjing University of Aeronautics and Astronautics, Nanjing 210016, China.*

[3]*Department of Mechanical and Automation Engineering, The Chinese University of Hong Kong, Shatin, Hong Kong.*

*Corresponding author: 0621072@zju.edu.cn

## Abstract

Two-dimensional mono-elemental material is an excellent saturable absorber candidate with low saturation intensity, large modulation depth, high nonlinearities, and fast recovery time of excited carriers. Typically, these mono-elemental material with two-dimensional structure possesses tunable bandgap from metallic to semiconducting according to different number of layers. The successful application of these materials as the saturable absorber has exploited the development of mode-locked fiber lasers. Therefore, this review is intended to provide an up-to-date information to the development of mono-elemental saturable absorber for the advances in mode-locked fiber laser, with emphasis on their material properties, synthesis process and material characterization. Meanwhile, issues and challenges of the review research topic will be highlighted and addressed with several concrete recommendations.

## 1.0     Introduction

Since the first passively mode-locked Nd:glass laser was demonstrated with a saturable absorber (SA) by De Maria et al. in 1966 [1], mode-locked laser was extensively investigated to generate ultrashort pulses in picosecond and femtosecond region. The intriguing study of ultrashort pulses have become a spotlight in photonics-related research due to its wide range of applications such as optical communication, biomedicine and diagnoses, material processing, spectroscopy and radar systems [2][3][4]. There are two main mechanisms to generate mode-locked laser either by active or passive approaches. In actively mode-locking, an external device such as function generator is required to drive the modulating signal of the modulator to generate mode-locked laser pulses [5]. An advantage of actively modulating approach is its feasibility to achieve higher repetition rate at higher harmonic orders by synchronizing the mode-locked laser with the modulating frequency. However, actively mode-locked laser shows limitations in weaker compactness, versatility, and heat dissipation issues compared to passively mode-locked laser [6].

The passively mode-locked laser is a well-known scheme by integrating a SA inside a laser resonator [7]. Under the irradiation of intense light, the absorption of the SA will be abruptly decreased which contributes to an increase in the transmission [8]. When the intense light interacts with a nonlinear optical (NLO) material, the electrons lying beneath the valence band tends to absorb incident photons and excited to the conduction band. Subsequently, saturable absorption occurs due to the Pauli-blocking mechanism, whereas a fully occupied conduction band could not further accept any excited photons [9]. Through saturable absorption, the SA periodically modulates the intra-cavity loss and transform free-running continuous wave (CW) laser into the ultrashort-pulsed laser. The CW laser is generated through stimulated emission, whilst a section of active gain medium is pumped by a laser source in a laser resonator. When the NLO parameters of the SA in a laser resonator is finely tuned, an array of pulse laser types could be realized as a consequence of the interaction dynamics between the intensity-dependent loss of the SA and the active gain medium [3].

The SA has been extensively proposed with several types of two-dimensional (2D) nanomaterials, whereas one of the earliest passively mode-locked fiber laser with 2D nanomaterial was demonstrated with graphene saturable absorber in 2009 [10]. Graphene is the most classic 2D nanomaterial which shows remarkable properties such as zero bandgap for broadband operation with wavelength range from ultraviolet to terahertz, ultrafast recovery time, electron mobility up to $10^6$ cm$^2$ V$^{-1}$ S$^{-1}$, and strong nonlinear refractive index of ~$10^{-7}$ cm$^2$/W [11][12][13]. However, several limitations of graphene is its low on/off capability in electronics structure and low absorption such as 2.3% of incident light per layer [10]. These issues restrict the integration of graphene for electronic nano-devices, meanwhile weaken its light modulation ability where strong light-matter interaction is required [14]. For the 2D materials composing of monolayer or few layer thicknesses, the strong light-matter interaction provides interesting study for the electronics and optoelectronics properties, such as the quantum-confined electronic band structure [15], bandgap tunability with different material thicknesses [16], and conductance transition with interplay between semiconductor and metal due to atom/lattice mechanism [17].

In comparison to graphene, mono-elemental group-V material in periodic table with 2D monolayer structure shows excellent electrical connectivity [18], high carrier mobility [19], topological transition states and negative Poisson's ratio [20], as well as tunable bandgap in conjunction to different thicknesses or material layers [21][22]. The 2D configuration of these mono-elemental group-V materials include phosphorene, arsenene, antimonene, and bismuthene. Phosphorene is the earliest studied 2D monolayers whose bandgap could be engineered such as 1.51 eV for black phosphorene [23] and 2 eV for blue phosphorene [24]. Moreover, arsenene and antimonene with wide bandgap were theoretically calculated instead of experimentally demonstrated [21]. Next, bismuthene was found to be topological non-trivial with large spin-orbit-coupling gap [25][26]. These mono-elemental materials have typical characteristics of low saturation intensity, large modulation depth, large third-order nonlinear optical response, broadband and tunable optical absorption, large optical damage threshold, and fast recovery time of excited carriers which are very suitable to be a good SA [27][28][29].

Recently, numerous reviews were done on synthesis technique and saturable absorption in 2D nanomaterials [8][30], transition metal dichalcogenides [31], black phosphorus [32][33], MXene [34], carbon allotropes (nanotube and graphene) and other 1D and 2D materials in ultrafast photonics and nonlinear optics [27][35]. However, a review on mono-elemental materials in mode-locked fiber laser remains at a rather infancy level. For instance, mono-elemental group-IV materials silicene and germanene were recently and firstly reported as saturable absorber for ultrafast fiber laser by Liu et al. in November 2020 [36] and Mu et al. in January 2021 [37], respectively. In comparison to group-IV material, earlier and wider demonstration of group-V material was reported as saturable absorber in year 2017 when black phosphorus was firstly made into its 2D form as phosphorene in mode-locked fiber laser [38]. In this review, the recent advances of the mono-elemental group-V materials in terms of its material properties, synthesis and material characterization, integration for SA fabrication and NLO investigation as well as its application in mode-locked fiber laser will be reviewed and studied. In addition, issues and challenges of this research study will be concluded and resolved with recommendations.

## 2.0    Material properties, synthesis and characterization of group-V monolayers

The post-nitrogen element in group-V exhibits thermodynamically stable monolayers by isolating the freestanding nanosheets from their layered allotropes [39]. These monolayer elements have five valence electrons in the valence shell orbital with the electronic configuration of $ns^2np^3$, whereby n is the number of inner shells to form a complete σ-system [40]. In the σ-system of a $sp^3$ hybridized element, strong in-plane chemical bonds and weak interaction are formed between the constitute atoms and between the layers of the material, respectively. Moreover, the in-plane covalent bonds are formed by three $sp^3$ hybridized orbitals, whilst the fourth $sp^3$ orbitals are constructed from non-bonding lone-pair electrons. These chemical bonding renders the group-V element with excellent semiconducting properties, electric conductance and thermal conductance, except that bismuth appears metallic in nature and shows non-trivial topological states [41]. Based on the study, the most stable regime of the group-V monolayer elements is puckered (α-phase) for phosphorene and buckled (β-phase) for the arsenene, antimonene and bismuthene [30].

### 2.1    Phosphorene

Phosphorus, a group-V element in the periodic table has the atomic number of 15 and atomic weight of 30.97. In nature, phosphorus appears as black, blue, violet, and several other amorphous forms [42]. Black phosphorus (BP) is the most prevalently studied material among these allotropes, which is constructed from strong in-plane covalent bonding network consisting of sp$^3$-hybridized phosphorus atoms and Van der Waals' force interaction bonding between its interlayer [43][44]. This unique chemical bonding structure contributes to the tunable direct bandgap for BP from ~0.3 eV for bulk BP to ~2 eV for monolayer BP (phosphorene) as shown in Fig. 1(a) and Fig. 1(b), respectively [22][45]. The BP and phosphorene has been widely employed as a potential SA candidate for the generation of ultrashort pulse owing to its tunable linear and nonlinear optical properties by the film thickness [46], higher absorption than monolayer graphene with 2.8% per BP layer [47], and strong third order NLO response, e.g. nonlinear refractive index of 10$^{-5}$ cm$^2$/W and third order nonlinear susceptibility of 10$^{-8}$ e.s.u [48]. Nonetheless, the environmental instability of the BP and phosphorene is its limitation to maintain high purity as the SA [48][49]. A solution to the oxidation issue of the BP is by implementing external protection such as dispersing the phosphorene quantum dot, an nanoparticle crystal of phosphorene into N-methyl-2-pyrrolidone (NMP), which was proven with excellent environmental stability for 6 months without degradation [38]. In addition, the implementation of graphene/phosphorene nano-heterostructure was proposed to protect the phosphorene from oxidation problem which also contributes to better performance than pristine graphene or phosphorene [50].

The 2D layered black phosphorus or phosphorene was firstly discovered for the application of field-effect transistor in 2014 [51]. Since then, numerous simulation and experimental works were done for the phosphorene. The phosphorene appears in puckered 2D monolayer structure of P$_6$ rings with a chair form. In this structure, each phosphorus atom is connected to three neighbouring atoms by covalent bonding with two bond lengths of 2.224 Å and 2.244 Å, as well as two bond angles of 96.34° and 103.09°, respectively as shown in Fig. 1(c) [30]. The $I_1$ bond connects the puckered sheet of the in-plane P atoms, whereas the out-plane P atoms are connected by $I_2$ bond. Subsequently, phosphorene exhibits strong in-plane anisotropy owing to its various structural parameters. For instance, the properties of the phosphorene are determined by the differences along the two axes (x-axis or y-axis) either the atoms are along the armchair or zigzag directions as presented in the inset of Fig. 1(c) [52]. For instance, the lattice constants along the armchair or zigzag directions are 3.30 Å and 4.53 Å, respectively [53]. The phosphorene along the armchair direction shows puckered structure whereas for zigzag direction, it appears as bilayer configuration as shown in Fig. 1(d) [53]. The interlayer distance between the layers is between 3.21 Å to 3.73 Å. Besides that, other remarkable properties of phosphorene include high carrier mobility of 1000 cm$^2$V$^{-1}$S$^{-1}$, on-off ratio of 10$^5$, high in-plane anisotropy, mechanical flexibility, and broadband spectral range demonstrated from 400 to 1930 nm [30][51][54][55][56].

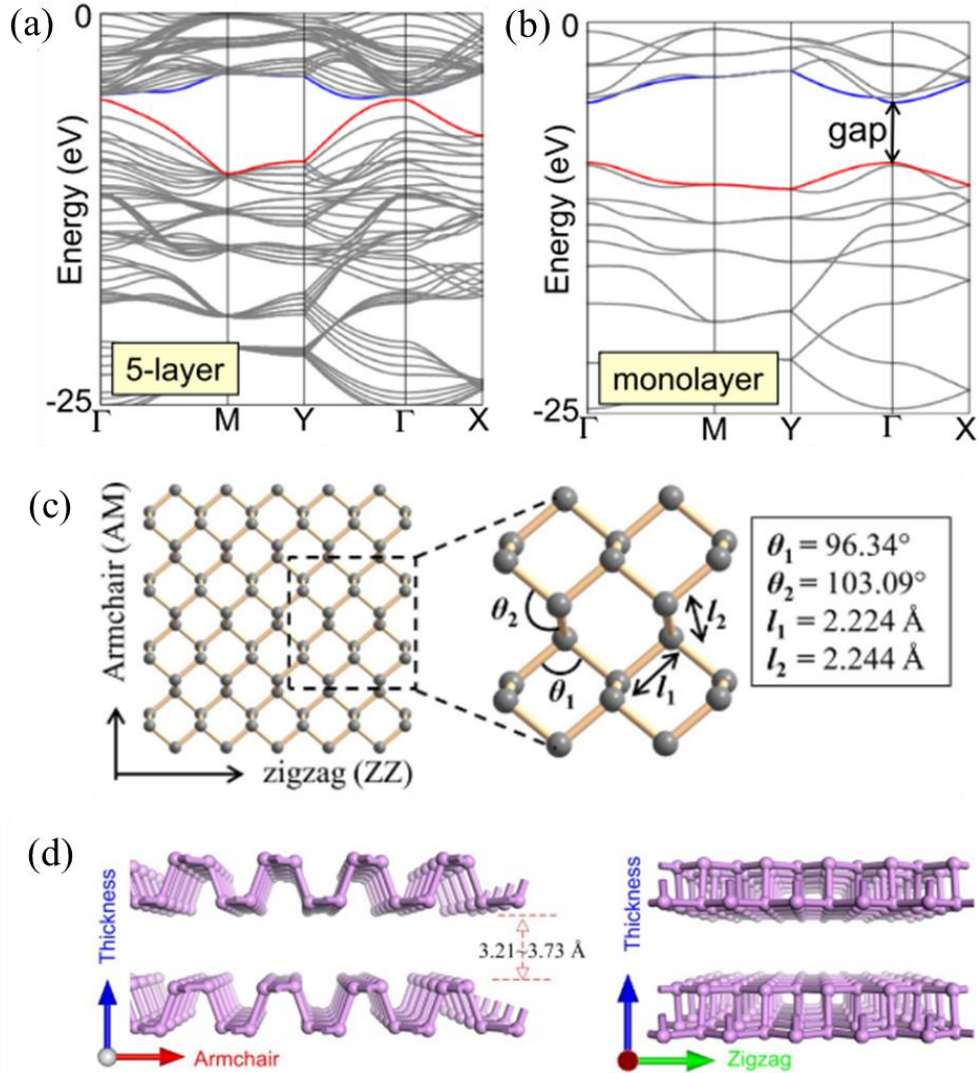

Fig. 1: Band structures for (a) bulk BP with bandgap of ~0.3 eV and (b) monolayer phosphorene with bandgap of ~2 eV, S. Fukuoka et al. [22]. © The Physical Society of Japan 2015. (c) Top view for atomic structure of P-P bonding for phosphorene, inset showing its structural parameters, P. Vishnoi et al. [30]. © Wiley-VCH Verlag GmbH & Co. KGaA, Weinheim 2019. (d) Side view for the armchair and zigzag direction, L. Kou et al. [53]. © American Chemical Society 2019.

The synthesis of phosphorene could be conducted by mechanical exfoliation [54]. The BP flakes were exfoliated repeatedly until monolayer phosphorene with ~0.85 nm of step height was attained using scotch-tape. The atomic force microscope (AFM) image of this monolayer phosphorene with ~0.85 nm step height was measured on a 300 nm $SiO_2$ substrate at the crystal edge as illustrated in Fig. 2(a). There is no photoluminescence signal for the bulk BP, whereas the phosphorene has a direct bandgap of 1.45 eV with a ~100 meV narrow width based on the photoluminescence measurement as shown in Fig. 2(b) [57]. This denotes that monolayer phosphorene has larger bandgap than bulk BP. However, the limitation of mechanical exfoliation is the difficulties for large scale fabrication of phosphorene and the resultant products inevitably consist of chemical residues from the scotch-tapes [54].

On the other hand, large-scale fabrication of phosphorene from bulk BP is feasible with liquid-phase exfoliation (LPE) method [58]. There are three main steps in the LPE method to form the colloidal solution of phosphorene, which are dissolve the bulk BP crystal in solvent, followed by sonication such as bath sonication and probe sonication, and finally the exfoliated BP sheet was separated from the non-exfoliated bulk via centrifugation. Some commonly used solvents include NMP [50][59][60], dimethylformamide (DMF) [61], N-cyclohexyl-2-pyrrolidone (CHP) [62], and isopropyl alcohol (IPA) [63]. Besides that, the phosphorene quantum dot with lateral size of 2.6 ± 0.9 nm was successfully synthesized through LPE method [38]. This phosphorene QD dispersion was characterized through a high resolution transmission electron microscopy (TEM) and the distance between adjacent hexagonal lattice fringes was measured to be 0.19 nm, which is close to the lattice spaces of (022) plane as shown in Fig. 2(c) [38]. The lattice spaces of the phosphorene were measured with selected-area electron diffraction (SAED) pattern which indicates the experimentally measured d-spacing of several planes was measured as ~0.1994 nm for (022) plane [64]. Based on [59], the d-spacing was measured as 0.323 nm for (012) plane and 0.224 nm for (014) plane, that indicates the phosphorene retains the crystalline state. More d-spacing of (111), (020), (121), (024), and (117) could be referred to the SAED pattern with 0.26 nm, 0.22 nm, 0.18 nm, 0.17 nm, and 0.13 nm, respectively in [60]. The Raman spectrum of the bulk BP and phosphorene QD is shown in Fig. 2(d). The 364.3 $cm^{-1}$, 440.5 $cm^{-1}$ and 467.1 $cm^{-1}$ Raman peaks are the characteristics of $A_g^1$, $B_{2g}$, and $A_g^2$ vibration modes of the phosphorene QD, respectively that represents different crystalline orientations within the layer plane [38]. The blue shift of the vibration peaks of ~4 $cm^{-1}$ to the bulk BP denotes that the phosphorene QD is either monolayer or bilayer [59]. Among these vibration modes, $A_g^2$ is most sensitive to the number of layers. Despite of the large-scale fabrication benefit, a disadvantage of LPE technique is the existence of unwanted defects brought upon by the time consuming bath sonication process [65].

Apart from the LPE method, the mass production of phosphorene could be done through electrochemical exfoliation [66][67][68][69]. This method employs the BP powder, stainless steel, and solvent as the raw material, electrode, and electrolyte, respectively. Some previously proposed solvents are deionized water [66], $Na_2SO_4$ [65] and aqueous $H_2SO_4$ solution [69]. By applying a potential difference, the electrolyte intercalates between the layers of BP electrode leading to the expansion and thus exfoliation into phosphorene. The utilization of different potential difference and electrolyte could result in different thickness of the exfoliated phosphorene, which was then separated from the electrolyte solvent by centrifugation. Subsequently, the X-ray photoelectron spectroscopy (XPS) spectra of bulk BP and electrochemical exfoliated BP is presented in Fig. 2(e). The $P2p_{1/2}$ and $P2p_{3/2}$ at the binding energy of 130.8 eV and 129.8 eV denote the crystalline BP peaks [66]. In addition, the broad peaks observed at around 133 eV binding energy indicates the oxygen bond ($P_{ox}$) of both bulk and exfoliated BP, whereas the exfoliated BP has more intense peak owing to its large surface phosphorene sheet that tends to be more sensitive to the surrounding environment, such as the presence of oxygen [66]. The absorption of monolayer to few-layer phosphorene was conducted through UV-VIS-NIR absorption spectrum as shown in Fig. 2(f) [65]. Based on the spectrum, one to five layers of phosphorene has strong absorption at 855 nm, 965 nm, 1160 nm, 1415 nm, and 1540 nm, which corresponds to the light energy of 1.45 eV, 1.29 eV, 1.07 eV, 0.88 eV, and 0.81 eV, in agreement with the photoluminescence measurement as previously mentioned in [54]. An advantage of electrochemical exfoliation is the feasibility for high yield product at different thickness and flake size [65].

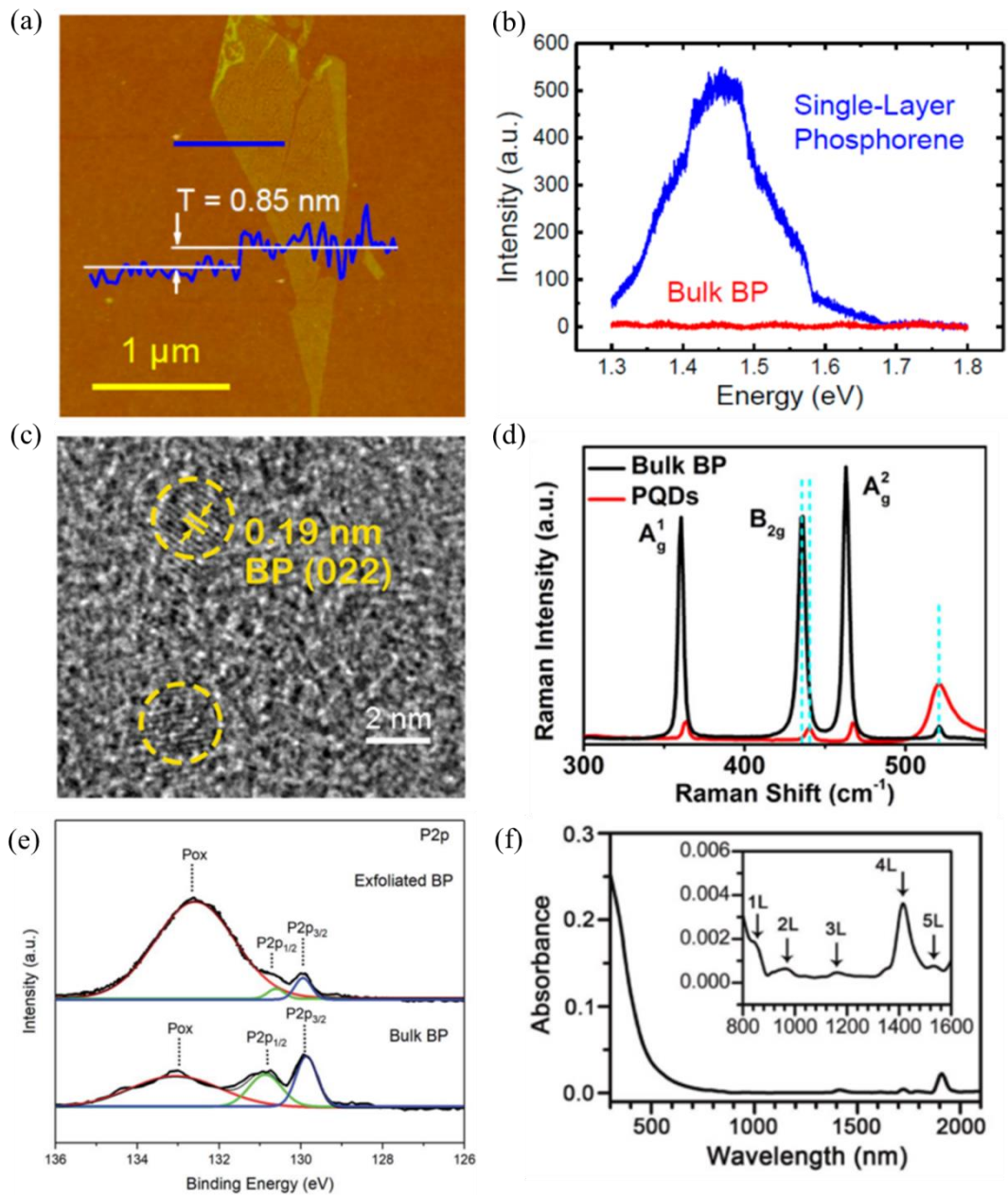

Fig. 2: (a) AFM image of the monolayer phosphorene crystal and (b) photoluminescence spectra of monolayer phosphorene and bulk BP on a SiO$_2$

substrate, H. Liu et al. [54]. © American Chemical Society 2014. (c) TEM image and (d) Raman spectrum for bulk BP and exfoliated BP, J. Du et al. [38]. © Creative Commons Attribution 4.0 International License 2017. (e) XPS spectrum for exfoliated and bulk BP, A. R. Baboukani et al. [66]. © The Royal Society of Chemistry 2019. (f) Absorption spectrum for 1 to 5 layers of phosphorene, J. Zheng et al. [65]. © Wiley-VCH Verlag GmbH & Co. KGaA, Weinheim 2017.

## 2.2    Arsenene

Arsenic is another group-V element in the periodic table with atomic number of 33 and atomic weight of 74.92. According to a density-function-theory calculation for arsenic, the bandgap for trilayer, bilayer, and monolayer of this element was deduced as 0 eV, 0.37 eV and 2.49 eV, respectively as shown in Fig. 3(a) [21]. Therefore, arsenic shows the feasibility for tunable bandgap from metallic properties to semi-conducting properties when the arsenic is engineered to be monolayer [70]. The monolayer arsenene was predicted with several structures, and the more preferable energetically which could withstand strong compressive and tensile strength is the buckled honeycomb structure, followed by puckered and lastly planar structures [71]. The atomic structures of the buckled, puckered and planar arsenene for side and top views are illustrated in Fig. 3(b). In likewise to phosphorene, the arsenic atom is connected to another three arsenic atoms via covalent bond with different bond angle and bond length for these three structures. Interestingly, arsenene with both buckled and puckered structures have indirect bandgaps of 1.635 eV and 0.831 eV respectively thus appearing as semiconductor, whilst planar arsenene is an unstable structure which shows metallic structure [71]. The indirect bandgap of the arsenene, especially with buckled and puckered structures could be tuned into direct bandgap by external tensile strain, as well as changing its material properties into topological insulator, which has quantum spin Hall state [72].

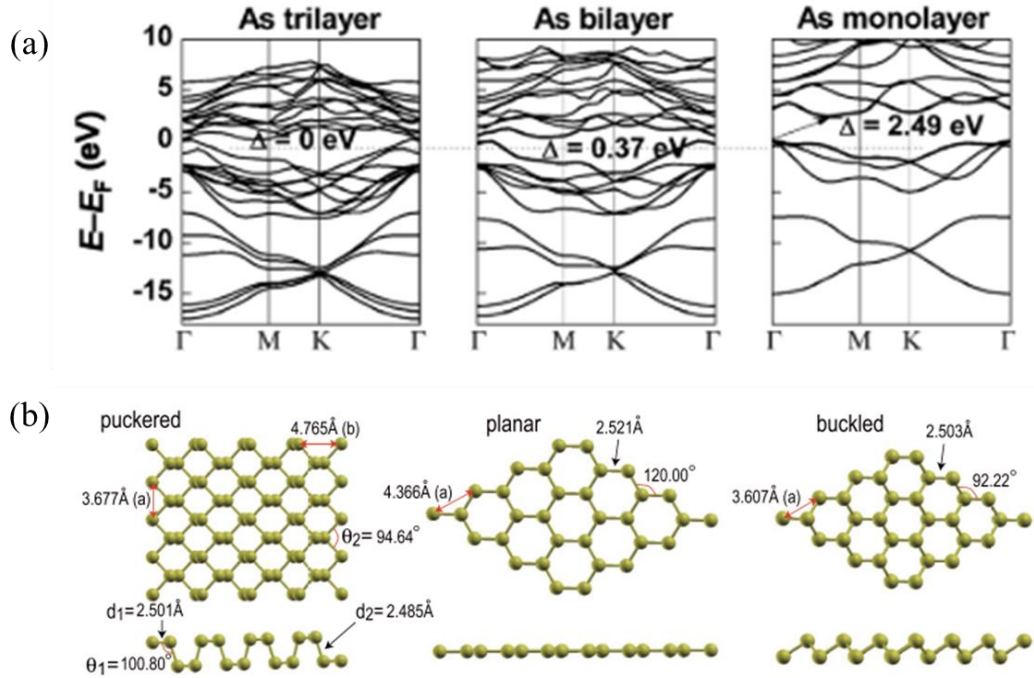

Fig. 3: (a) Band structures for 1, 2, and 3 arsenic layers, S. Zhang et al. [21]. © Wiley-VCH Verlag GmbH & Co. KGaA, Weinheim 2015. (b) Atomic structures for puckered, planar and buckled arsenene, C. Kamal et al. [71]. © American Physical Society 2015.

Before the latest demonstration for the synthesis of monolayer arsenene by Shah et al. [73], large experimental effort was contributed to fabricate multilayer arsenene in the recent few years. For instance, multilayer arsenene nanoribbons were fabricated on indium nitride (InN) and indium arsenide (InAs) by the plasma-assisted method [74]. During the process, ~14 nm of multilayer arsenene was fabricated on a ~13.8 nm InN and 5 nm InAs as shown in the TEM image in Fig. 4(a). The $SiO_2$ layer was capped on the arsenene surface to protect this thin film from damage due to the focus ion beam during the preparation of the TEM sample. The inset of Fig. 4(a) was mapped through fast Fourier Transform to derive the d-spacing of the multilayer arsenene. Based on the atomic model in Fig. 4(b), the d-spacing of the (100) and (01-1) planes were derived as ~0.277 nm and ~0.188 nm, respectively. Subsequently, the arsenic atoms were separated from the InAs surface to form multilayer arsenene nanoribbons with an estimated bandgap of ~2.3 eV, owing to the green light emission at ~540 nm from the photoluminescence measurement which matches with the calculated value after annealing [74]. Apart from the plasma-assisted method, aqueous shear exfoliation was employed to prepare high-yield arsenene nanosheet [75]. The arsenene nanosheet was exfoliated from raw material, grey arsenic which has rhombohedral layered structure thus contributing to its very brittle and easily pulverized properties [76]. There are three main steps for the aqueous shear exfoliation process. The pre-treatment of bulk arsenic is the first step of the process via purifying the powder and removal of oxides on the arsenic crystal surface. Sonication of arsenic powder took place in aqueous surfactant sodium cholate in an ice bath and the suspension was then centrifuged, leaving the powder to be dried in a vacuum oven at 60 °C. The second step of the process is the shear dispersion and exfoliation of the pre-treated powder by shear mixers with opposite rotational direction. The third step is the centrifugation of the suspension, in which different thicknesses of arsenene nanosheet was finally attained by aqueous washing and centrifugation at different rotational speed.

Another synthesis process for 6-12 layers of arsenene nanosheets were presented by probe sonication of grey arsenic in NMP under inert atmosphere [77]. Firstly, 20 mg of grey arsenic was added with 20 mL of de-gassed NMP under a nitrogen atmosphere. The probe containing the mixture was then sonicated with an input intensity of 200 W for 12 hours. After the sonication process, the black dispersion was centrifuged at 1000 rpm for 30 minutes and the supernatant containing few layer arsenene was collected as the final product for further characterization. The Raman spectrum of the grey arsenic, arsenene nanosheet and nanodots is shown in Fig. 4(c). The Raman characteristics peaks of ~196.1 cm$^{-1}$ and ~258.5 cm$^{-1}$ agree with the in-plane vibration and out-of-plane vibration modes for $E_g$ and $A_{1g}$ of the arsenic allotropes [78]. In addition, the XPS spectra for few-layer 3d and 3p arsenene is shown in Fig. 4(d). For the 3d arsenene, the XPS spectrum exhibits a broadband covering from ~40 to 43 eV, which are assigned as the $3d_{3/2}$ and $3d_{5/2}$ bands of the arsenic [79]. Moreover, the XPS spectrum for the 3p arsenene shows the $3p_{1/2}$ and $3p_{3/2}$ bands for the arsenic, which corresponds to the binding energy of ~145 eV and ~140 eV. This is in accordance with the previous measurement that the $3p_{1/2}$ has higher binding energy than $3p_{3/2}$ bands for the arsenic [80]. Recently, a monolayer arsenene was successfully demonstrated with average lattice constant of 3.6 Å, which is consistent with buckled arsenene derived from low-energy electron diffraction (LEED) [73]. The monolayer arsenene was synthesized by exposing the arsenene on Ag (111) substrate to an arsenic pressure of $2 \times 10^{-7}$ Torr, at a temperature of 250 °C to 350 °C for 3 minutes. Subsequently, the monolayer arsenene was atomically resolved along the black line in the selected area of a scanning tunnelling microscopic (STM) image as shown in the inset of Fig. 4(e) and the value of lattice constant was then deduced from the topographic profile as shown in Fig. 4(f).

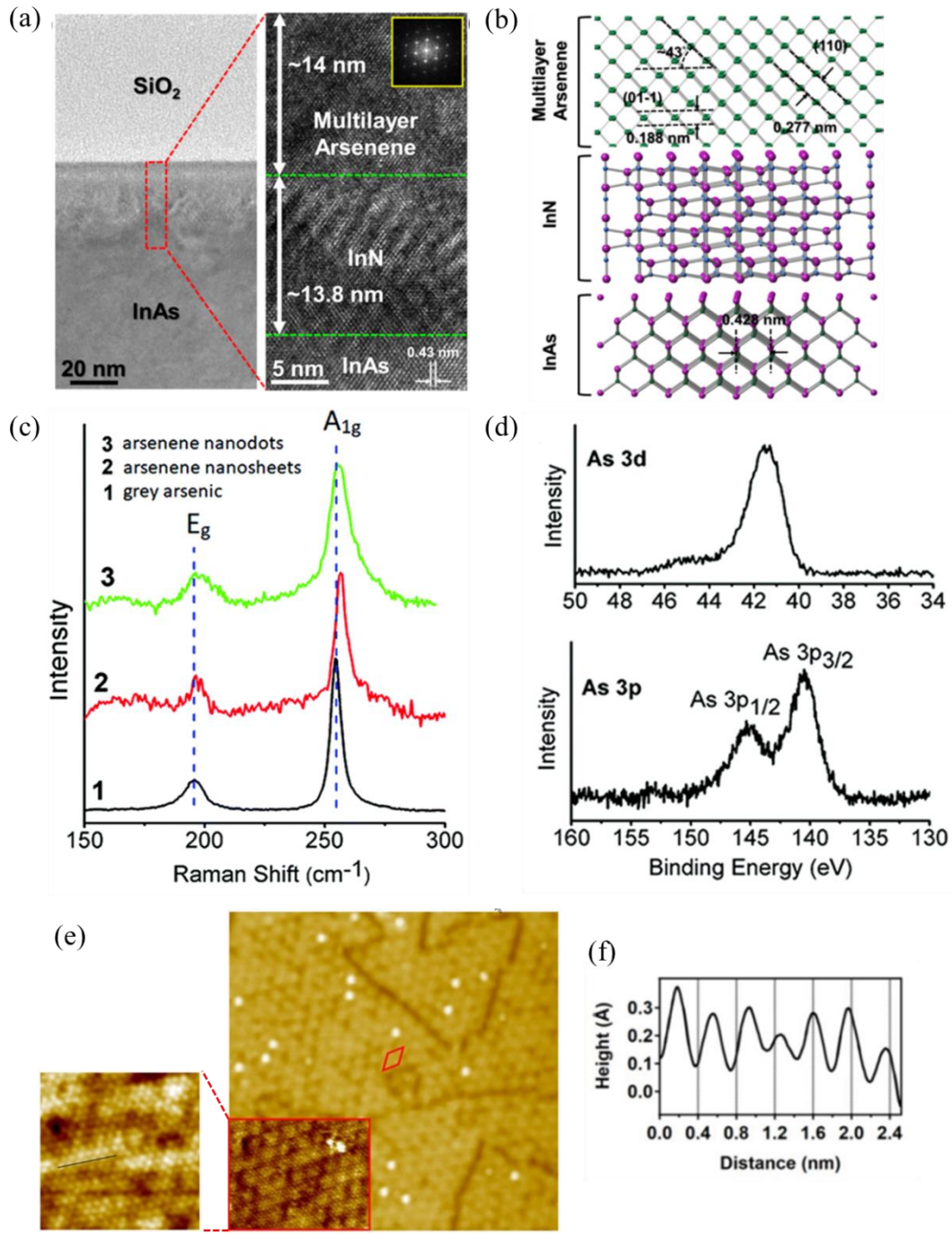

Figure 4: (a) TEM image and (b) theoretical atomic model of the multilayer, InN, InAs structure, H. Tsai et al. [74]. © American Chemical Society 2016. (c) Raman spectrum and (d) XPS spectrum for the grey arsenic, arsenene nanosheet and nanodots, P. Vishnoi et al. [77]. © The Royal Society of Chemistry 2018. (e) STM image and (f) topographic profile for monolayer arsenene, J. Shah et al. [73]. © IOP Publishing Ltd 2020.

## 2.3 Antimonene

Antimony is the third group-V element in the periodic table which has atomic number of 51 and atomic weight of 121.76. In likewise to arsenic, the bandgap of the antimony was predicted with the density-function-theory calculation. Based on the calculation, monolayer antimonene was predicted with an indirect bandgap of 2.28 eV, whereas its bilayer and thicker structure exhibits metallic properties without a bandgap [21]. The thickness of monolayer antimonene is about 0.373 nm [81]. The band diagram structure of the monolayer, bilayer, and trilayer antimonene is shown in Fig. 5(a). The antimonene has similar characteristic to arsenene in terms of its tunability of indirect bandgap into direct bandgap for its monolayer structure under biaxial strain [82][83]. The buckled antimonene monolayer possesses quantum Spin Hall effect by inhibiting mutual hybridization between the in-plane and out-of-plane orbitals [84]. This structure could become a graphene-like flat honeycomb when the in-plane $sp^2$ does not hybridize with the out-of-plane $p_z$ orbitals. The monolayer antimonene consists of two phases, α- and β-antimonene. The α-antimonene has strong anisotropic characteristic, but it is opposite for the β-antimonene which has nearly isotropic characteristic within the monolayer plane [85]. The atomic structure of the α- and β-antimonene is shown in Fig. 5(b). The interlayer distances between the layers of α- and β-antimonene are 6.16 Å and 3.65 Å, respectively. Xu et al. [86] had investigated the electronics and optical properties of α- and β-allotropes of monolayer antimonene. Based on their investigation, the α-antimonene possesses smaller direct bandgap than the indirect bandgap of their β-counterpart. Nevertheless, the α-antimonene is an excellent candidate for saturable absorber due to its excellent absorption properties. The NLO response of an antimonene dispersion was examined by spatial self-phase modulation (SSPM) at 1064 nm of wavelength, proving its effective nonlinear refractive index of ~$10^{-5}$ $cm^2W^{-1}$, third-order susceptibility of ~$10^{-8}$ e.s.u. and relative change of the nonlinear refractive index ranging from 14% to 63% at different intensities [87]. In contrast to phosphorene, antimonene with high environmental stability and antioxidant capacity is difficult to be oxidized in atmosphere at room temperature [83][88]. In the contrary, a disadvantage of the monolayer antimonene is its indirect bandgap that might give rise to low photoelectric response which causes challenges for the application in ultrafast photonics [89].

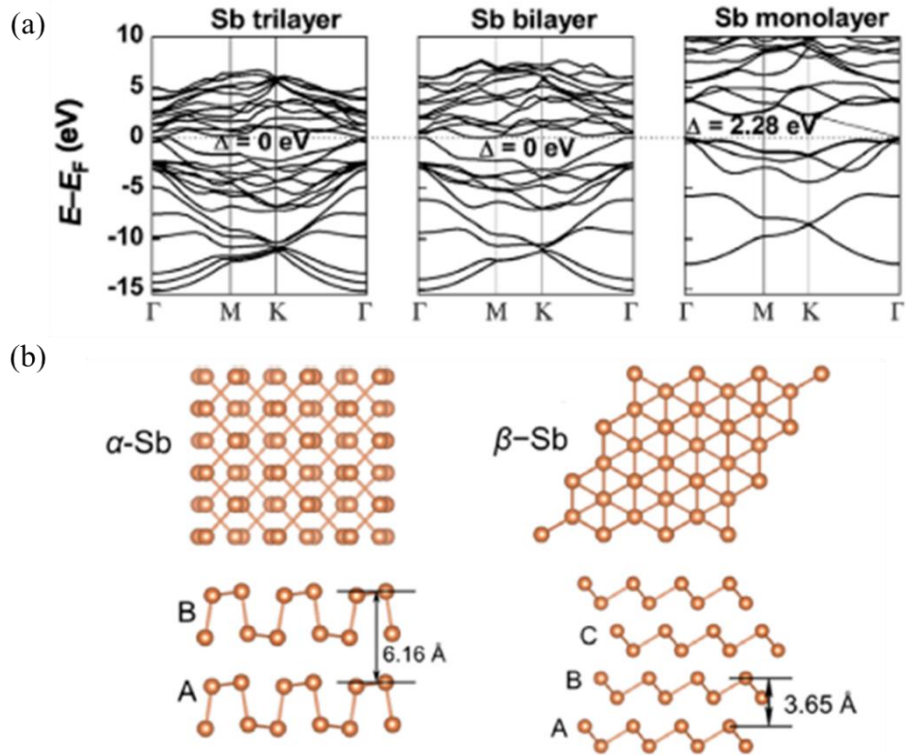

Fig. 5: (a) Band structures for 1, 2, and 3 layers antimonene, S. Zhang et al. [21]. © Wiley-VCH Verlag GmbH & Co. KGaA, Weinheim 2015. (b) Atomic structures for α- and β-antimonene, G. Wang et al. [85]. © American Chemical Society 2015.

Mechanical exfoliation is a method to synthesize free-standing, rhombohedral-structured monolayer β-antimonene with a thickness of closely 0.4 nm from bulk antimony crystals on $SiO_2$/Si substrates [81]. During the exfoliation, a viscoelastic stamp was integrated as an intermediate substrate to transfer the antimonene nanoflakes on the $SiO_2$/Si substrates. Fig. 6(a) shows the AFM image of the antimonene and its step profile along the green line. The lowest step height of the antimonene terrace measured from AFM at ~0.4 nm validates the monolayer structure of the antimonene with a bandgap of ~1.2 eV to 1.3 eV based on the DFT calculation [90]. However, the scalable production remains an issue for the mechanical exfoliation [91]. In this case, scalable β-monolayer antimonene with high quality could be synthesized via MBE method on several substrates, such as palladium telluride ($PdTe_2$) [92], Ge(111) [93] and Ag(111) [94]. In [92], the antimony atoms were evaporated from a Knudsen crucible and deposited onto a pre-cleaved $PdTe_2$ substrate at 400 K. Based on the low-energy electron diffraction and STM measurements, the thickness of the monolayer antimonene grown on $PdTe_2$ is ~2.8 Å as shown in Fig. 6(b) [92]. Fig. 6(c) illustrates the XPS core-level spectrum of the Sb 4d monolayer antimonene (orange line) with two peaks, $4d_{3/2}$ and $4d_{5/2}$ at 32.35 eV and 32.05 eV, respectively. The violet line denotes the small antimony cluster on the substrate and the red line indicates the antimony in the antimonene. Despite the scalable fabrication advantage of the MBE method, a disadvantage of this method is its high dependency on close lattice matching. Subsequently, Van der Waals' epitaxy growth provides a viable option to synthesize layered materials with large lattice mismatch [95]. This method provides the growth of a layered material onto another layered material without dangling bond on its cleaved surface. For instance, β-antimonene polygons with lateral size of ~5 to 10 μm and thickness of ~4 nm were grown on a fluorophlogopite mica substrate by Van der Waals' epitaxy in a two-zone furnace [96]. The Raman spectrum for the bulk antimony and various thickness of antimonene is shown in Fig. 6(d). The Raman peaks at 110.7 cm$^{-1}$ and 149.2 cm$^{-1}$ for the bulk antimony indicates the $E_g$ and $A_{1g}$ vibration modes [97]. In addition, the Raman peaks were blue-shifted to a higher wavenumber region by decreasing the sheet thickness, which could be due to the lattice constant shrinks and long-range Coulombic interlayer interactions [98][99].

Electrochemical exfoliation was conducted to exfoliate antimony crystal into ultrathin film using a 0.5 mm diameter of platinum wire as the anode and sodium sulphide ($Na_2SO_4$) as the electrolyte [100]. The $Na^+$ ions released from $Na_2SO_4$ electrolyte intercalates the antimony crystals and facilitates the exfoliation process. With a fixed potential of -6V, the electrochemical process was carried out for an hour before the antimonene nanosheets were centrifuged at 6000 rpm for 30 minutes. As a result, the nanosheets with a thickness of ~31.66 nm and lateral dimension of ~10.3 μm were synthesized as the final product. Moreover, simultaneous electrochemical exfoliation and fluorine functionalization was carried out in a three-electrodes electrochemical cell using [BMIM][$PF_6$]/$CH_3CN$ solution as the electrolyte [101]. Fig. 6(e) shows the hexagonal profile of few layer antimonene and its atomic structure is in analogy with the crystal lattice as shown in the TEM image in Fig. 6(f). The interplanar spacing is 2.18 Å which is in good agreement with the (110) plane of the β-antimonene single crystal. In addition, the absorption of few layer antimonene dispersed in MeCN is increased in the NIR region as portrayed in Fig. 6(g). Therefore, this fluorine functionalized antimonene nanosheets show strong NLO response, which was demonstrated for the generation of passively Q-switching laser pulses in NIR region. Apart from electrochemical exfoliation, the LPE method was used to form the few-layer antimonene nanoflakes, in which the bulk antimony is firstly immersed in 1 mg/mL NMP, and 2D layered-material is obtained through ultrasonic exfoliation operating with a frequency of 40 kHz for 4 hours [83]. In another preparation of antimonene nanoflakes through the LPE method, the process was started with the grinding of 300 mg antimony crystals using agate mortar, the grinded antimony powder were then collected and dispersed in 6 mL ethanol for 2.5 hours with ultrasonic treatment using 40 kHz and 300 W at room temperature [102]. Next, the antimonene suspension was sealed for 48 hours and centrifuged for 30 minutes with rotation speed of 1500 rpm. Finally, the large antimonene particles were precipitated and leaving only the antimonene nanoflakes as the supernatant. In addition, the LPE was also employed together with solvothermal method in preparing antimonene dispersion [82]. In the process, the antimonene crystals were grinded into antimonene powder with ball-milling for 20 hours at 400 rpm and immersed in 0.5 mg/mL ethanol. After sonication of 6 hours, the obtained dispersion was then centrifuged at 400 rpm for 4 minutes. The supernatant was then heated up to 50 °C for 2 hours to isolate oxygen, then the solution was immediately added into a reaction kettle at 140 °C for 6 hours to produce the antimonene dispersion.

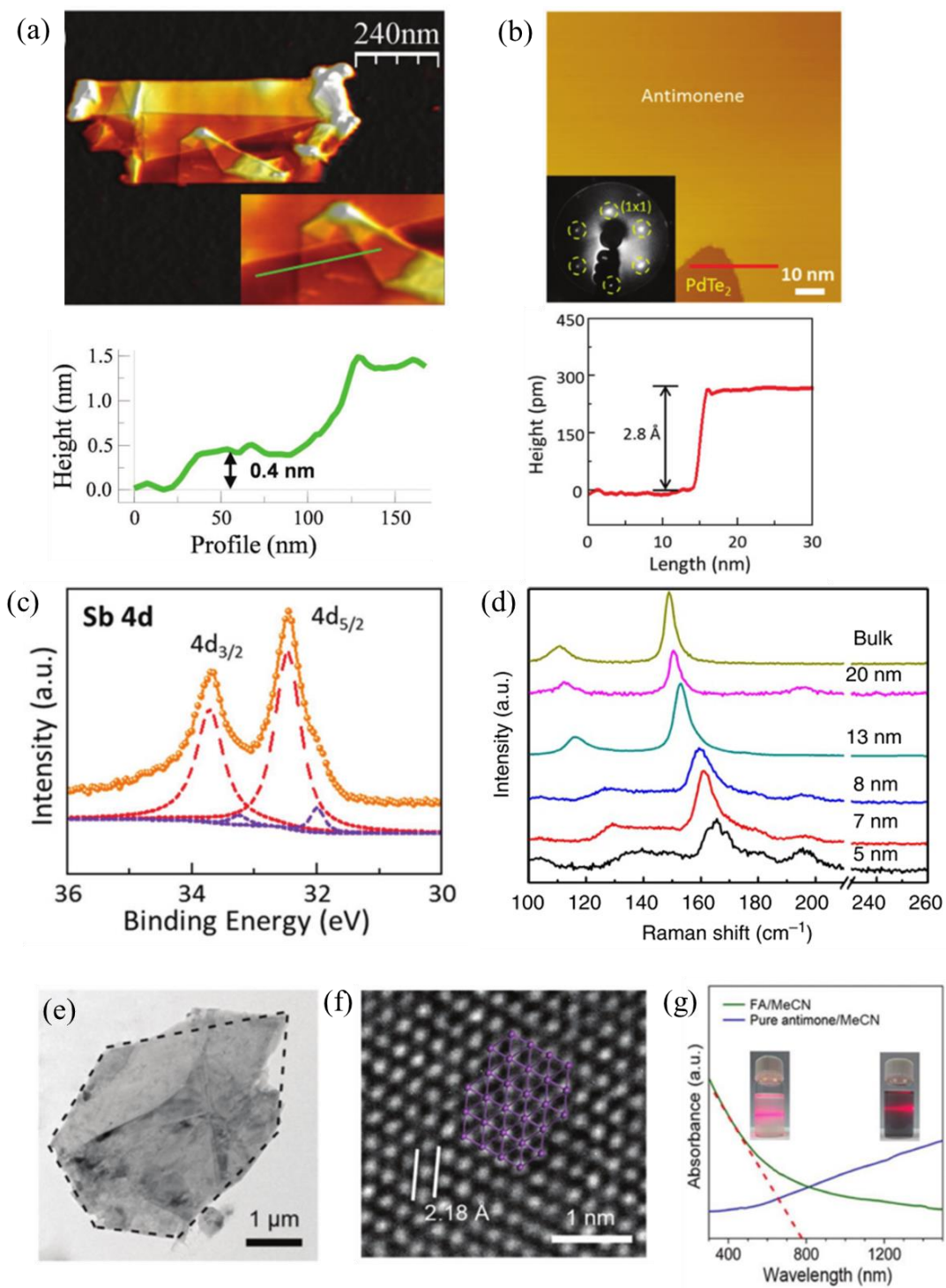

Fig. 6: (a) AFM image and its profile along the green line of the antimonene, P. Ares et al. [81]. © Wiley-VCH Verlag GmbH & Co. KGaA, Weinheim 2016. (b) STM topographic image of antimonene islands on PdTe$_2$ substrate and its height profile along the red line, (c) Sb 4d XPS core-level spectrum, X. Wu et al. [92]. © Wiley-VCH Verlag GmbH & Co. KGaA, Weinheim 2017. (d) Raman spectrum of bulk

antimony and few-layer antimonene with different thickness, J. Ji et al. [96]. © Creative Commons Attribution 4.0 International License 2017. (e) TEM image, (f) HRTEM image and (g) Absorption spectrum of few layer antimonene nanosheet dispersed in MeCN, G. Zhang et al. [101]. © The Royal Society of Chemistry 2019.

## 2.4 Bismuthene

Bismuth, the fourth group-V element in the periodic table has the atomic number of 83 and atomic weight of 208.98. Owing to its intrinsic spin-orbital coupling characteristic, bismuth facilitates the application of spintronic devices without the needs for external strong magnetic field [26][103]. In comparison to metals, the semi-metallic or semiconducting characteristics of bismuth render its investigation into condensed matter physics because of the long Fermi wavelength of its electronic properties [104][105]. For instance, bismuth shows feasibility in transition from semi-metal to semiconductor according to the quantum confinement effect by reducing the thickness of the material [106]. Bismuth in its 2D monolayer form, bismuthene is incredible for its quantum Hall effect which shows a similar honeycomb lattice with monolayer graphene, or β-phase of arsenene or antimonene [107]. The atomic structure of the bismuthene is shown in Fig. 7(a). Based on the structural arrangement, each bismuth atom is connected to three neighbouring bismuth atoms with bond angle of 93.27°, bond distance between bismuth atoms of 3.04 Å, buckling height of 1.73 Å, and lattice constant of 4.357 Å [108]. The interlayer distance between two bismuthene layers are 3.9 Å [109]. The monolayer bismuthene has the smallest bandgap among the group-V mono-elements. A study shows that bismuthene possesses a stable buckled hexagonal structure and layer-dependent (1 to 6 layers) energy gap from ~0 to 0.55 eV as shown in Fig. 7(b), thus it is suitable for broadband operation from terahertz, mid-infrared toward near infrared region, and even visible wavelength region [109][110]. In addition, monolayer bismuthene with 0.3 to 0.5 eV bandgap was proven to be topologically non-trivial, contributing to its significant values as 2D topological insulator [25][111]. The topological insulators have been extensively proposed as potential SA candidate to generate ultrashort pulses [112][113]. Zhang et al. [114] had proposed the allotropes of bismuthene in α- and β-phases as direct bandgaps of 0.36 eV and 0.99 eV, respectively. In addition, the puckered structured bismuthene could transform the indirect bandgap to direct bandgap with 0.18 eV and 0.23 eV for α- and β-phase, respectively [114]. An advantage of bismuthene is its stability to resist oxygenation in the atmosphere compared to the BP or phosphorene [115].

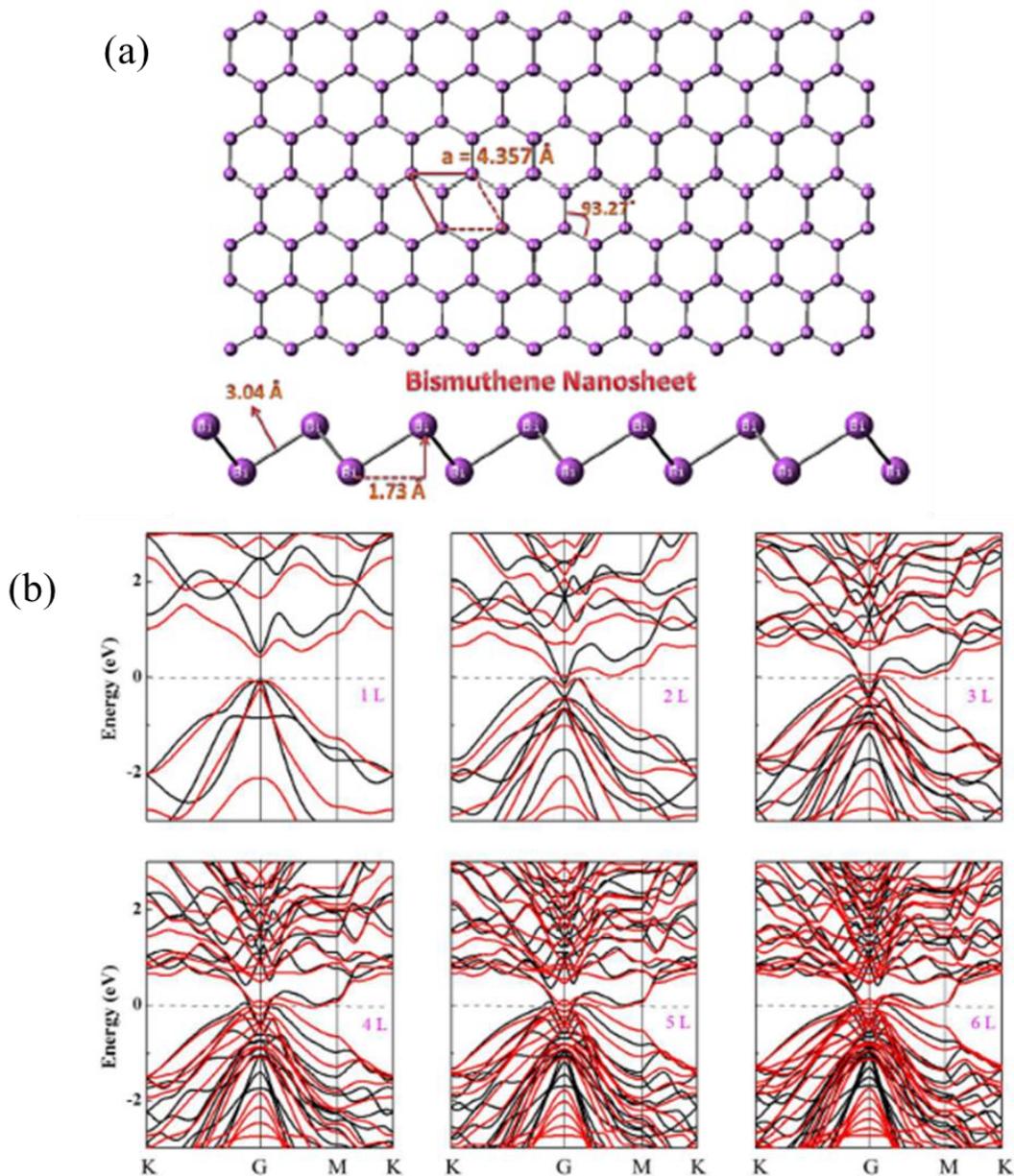

Fig. 7: (a) Atomic structures of bismuthene nanosheet, R. Bhuvaneswari et al. [108]. © Elsevier 2019. (b) Band structures for 1-6 layers bismuthene, L. Lu et al. [109]. © American Chemical Society 2017.

Molecular beam epitaxial (MBE) was carried out for the early synthesis of atomically thin bismuthene with ~1 μm edge on Si surface [116]. The MBE method was also implemented to prepare monolayer bismuthene on a silicon carbide (SiC) (0001) substrate that possesses two advantages, which are to provide stability to bismuthene as a quasi-2D topological insulator and to achieve larger tunable bandgap [117]. During this MBE process, the n-doped Si-terminated 4H-SiC substrate was firstly dry-etched in a hydro-based gas atmosphere at 1230 °C. After that, the H-termination of the substrate was removed by a slow thermal H desorption at 650 °C. At the same time, the bismuth atoms were immediately bombarded on this substrate. Finally, the substrate was treated with lower temperature of 500 °C to condense the bismuth layer, in which a honeycomb lattice monolayer bismuthene with a topological energy bandgap of ~0.8 eV was synthesized as the final product [118].

Apart from the MBE method, free-standing few-layer bismuthene was prepared through sonochemical exfoliation from bulk bismuth crystal [110]. In the preparation process, the bulk bismuth was firstly grinded into bismuth powder which was then mixed with isopropyl alcohol with a ratio of ~0.052:0.948. Next, the mixture was sealed in a spiral glass bottle and undergoes probe sonication and bath sonication for 10 hours. Finally, the suspension was centrifuged at 5000 rpm for 20 hours. The bismuth powder and bismuthene suspension were characterized in thorough. Fig. 8(a) shows the TEM image of the bismuthene, proving its lateral size of ~0.8 µm. The HRTEM in Fig. 8(b) shows the arrangement of lattice planes with interlayer distance of 0.322 nm, which corresponds to the (111) plane of rhombohedral A7 structure [119][120]. The crystalline structure of the bismuthene was proven with the SEM image in Fig. 8(c), whereas the bismuthene nanoflake has a lateral height of ~4 nm with a smooth surface and irregular profile as shown in the AFM image in Fig. 8(d). The nonlinear refractive index of the sonochemical-exfoliated bismuthene was measured as ~$10^{-6}$ $cm^2$/W and third-order susceptibility as ~$10^{-9}$ e.s.u. through spatial self-phase modulation, contributing to its potential in passively mode-locked laser [110]. Another sonochemical exfoliation method was reported to synthesize few-layer bismuthene dispersion [89]. In the process, the bulk bismuth crystal was firstly grinded into bismuth powder, which was then added into ethanol solution with a ratio of 0.03:0.967 in a glass bottle. Next, the glass bottle was placed under ice-bath sonication and probe sonication for 15 hours. Finally, the bismuthene dispersion is attained after the suspension was centrifuged for 20 minutes at 7000 rpm.

Moreover, few-layer bismuthene could also be synthesized through bath sonication and probe sonication from bulk bismuth in the suspension of ethanol or isopropanol [121][122]. The suspension was then centrifuged and the supernatant was extracted repeatedly. Based on the absorption measurement in Fig. 8(e), the highest absorption peak of the bismuthene was observed from 1300 nm to 1600 nm [122], particularly at 1550 nm thus makes this material very useful for the communication band application [121]. In addition, the high absorption at ~1800 to 2000 nm are very promising for the bismuthene as the SA. Fig. 8(f) and Fig. 8(g) shows the Raman spectrum of few-layer bismuthene and the comparison between the bulk bismuth and bismuthene, respectively. The Raman peaks at 69.3 $cm^{-1}$ and 97 $cm^{-1}$ indicate the $E_g$ and $A_{1g}$ vibration modes of the bismuthene [123]. According to Fig. 8(g), the Raman peak intensity of bismuthene is stronger than bulk bismuth because the electrons are tightly bound in the bulk bismuth [124].

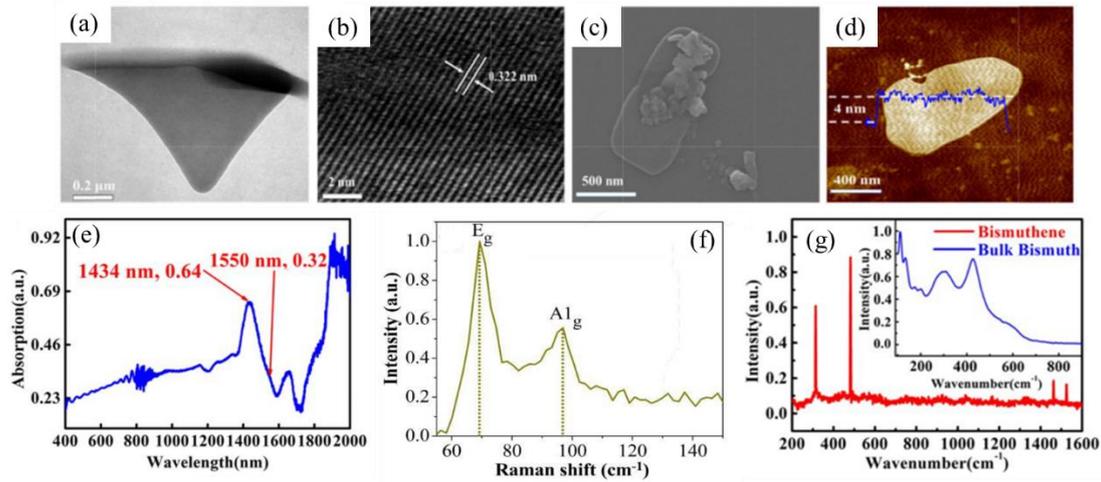

Fig. 8: (a) TEM, (b) HRTEM, (c) SEM, and (d) AFM images of bismuthene nanoflakes, L. Lu et al. [110]. © Wiley-VCH Verlag GmbH & Co. KGaA, Weinheim 2017. (e) Absorption spectrum of bismuthene, P. Guo et al. [121]. © IOP Publishing Ltd 2019. (f) Raman spectrum of bismuthene from ~60 to 150 cm$^{-1}$ wavenumbers, T. Feng et al. [125]. © Creative Commons Attribution 4.0 International License 2019. (g) Raman spectrum of bismuthene from ~60 to 150 cm$^{-1}$ wavenumbers, P. Guo et al. [121]. © IOP Publishing Ltd 2019.

## 3.0 Saturable absorber

The SA is an optical component which absorbs certain amount of light and its absorption is reduced at high intensities. The SA is important to generate the passively mode-locked fiber laser. In this section, the fabrication of the SA and their NLO characterization with both balance twin-detector and Z-scan measurement setups will be discussed.

### 3.1 Fabrication of saturable absorber

There are several optical fibre platforms for the fabrication of the SA by the interaction of the material and the propagated light signal along the optical fibre [126]. The easiest platform is the insertion of the SA material between the optical fiber ferrules, at the expense of short interaction length and low optical damage threshold [127][128][129]. For example, an optical fiber tip was dipped into the antimonene supernatant for 10 minutes, and drying for 5 hours to form the SA [102]. The 0.05 mL phosphorene SA was integrated with the optical fiber end facet for the NLO characterization [59]. Apart from fiber ferrule structured SA, a D-shaped fiber with interaction length of 10 mm, and distance from the fiber core boundary to the D-shaped area of 2 μm was immersed in the phosphorene/graphene solution to form the SA [50]. The D-shaped fiber SA has longer NLO interaction length and higher optical damage threshold than the fiber ferrule SA due to the indirect interaction with the material at its cleaved surface that induces evanescent field [127][130]. Apart from D-shaped fiber, the optical fibre which was tapered down into several micron sizes could also generate evanescent field around the waist region for the interaction with the SA materials [131][132]. Several advantages of the tapered fiber include the high optical power damage threshold, excellent heat dissipation mechanism and long NLO interaction length for the SA materials [133][134]. The tapered fiber could be fabricated using the flame taper method as shown in Fig. 9(a) [83][121]. In the tapering process, a piece of protective layer stripped single-mode optical fibre was stretched with heat by the flame sprayer to be tapered down into desired waist length and diameter. There are two approaches for the interaction of the waist region of the tapered fiber to the SA materials which are via the drop casting or optical deposition as shown in Fig. 9(b) and 9(c), respectively. The SA materials in solution or dispersion form could either be drop casted on the waist region of the optical fiber [125], or an additional source such as continuous wave (CW) laser could be employed for the optical deposition process and the overall process was monitored via an optical power meter [110].

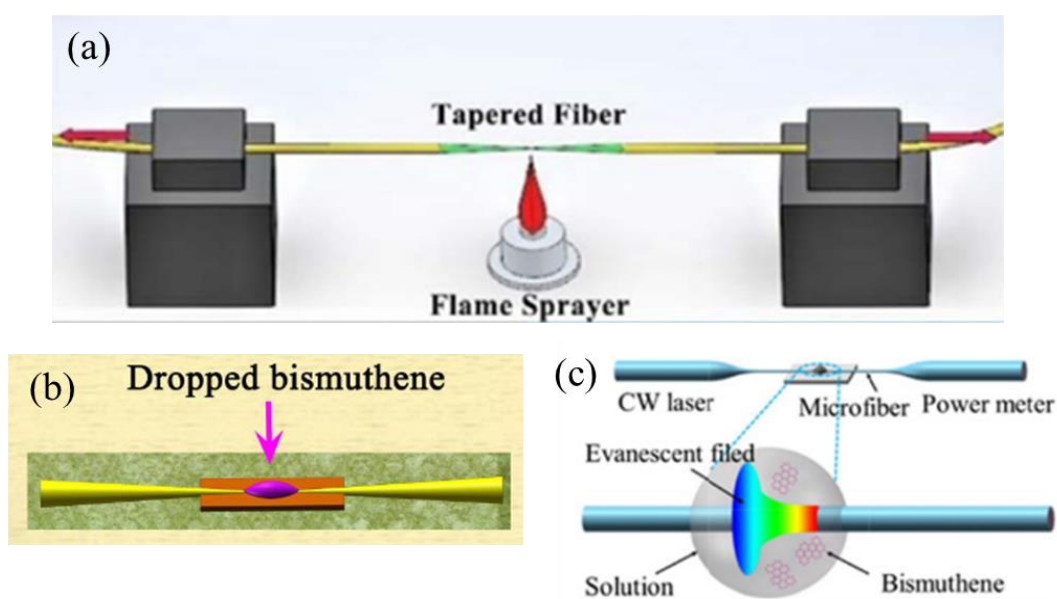

Fig. 9: (a) Flame taper method, P. Guo et al. [121]. © IOP Publishing Ltd 2019. (b) Drop casting, T. Feng et al. [125]. © Creative Commons Attribution 4.0 International License 2019. (c) Optical deposition of SA solution/dispersion at the waist region of the tapered fiber, L. Lu et al. [110]. © Wiley-VCH Verlag GmbH & Co. KGaA, Weinheim 2017.

Based on the critical review of the group-V mono-elements, tapered fiber is the most commonly utilized platform to fabricate the SA, which could be due to its technology maturity and the advantages of this platform compared to fiber ferrule or D-shaped fiber such as feasibility to withstand high optical power and insensitive to polarization-dependent loss, respectively. Nevertheless, the fragility of tapered fiber especially at its waist region which is easily exposed to environmental contamination is its main disadvantage [135]. A tapered fiber was used as the platform, whereas the phosphorene nanoflakes [65], bismuthene solution [136][110], and bismuthene/ethanol dispersion [122] were used to drop cast or fill the evanescent area of the tapered fiber to form the SA. The deposition length for the phosphorene nanoflakes is ~176 μm on a tapered fiber with waist diameter of ~4.4 μm [65]. The tapered parameter is measured at waist diameter of ~15 μm and tapered length of 1 mm in [122], as well as waist diameter of 16 μm and waist length of 5.5 mm in [136]. Apart from the drop casting method, Table 1 summarizes the dimension of the tapered fiber which was used to fabricate the SA through optical deposition technique with different types material solution or dispersion. The CW laser power was kept constant at 60 mW in [38], whereas an increase of CW laser power from 55 mW to 360 mW was utilized in [83]. The deposition length after the optical deposition is measured as ~240 μm [83] and ~1.5 mm [124], respectively. In [125], the optical deposition was carried out for 40 minutes with a decrease in optical power from 36.6 mW to 17 mW, which indicates that 54% of the power is absorbed by the bismuthene. The absorption value of ~54% was also reported in [115].

Table 1: Summary of optical deposition method to fabricate SA using Group-V 2D mono-elements on the tapered fiber.

| Input source | Waist length | Waist diameter | Material | Ref. |
|---|---|---|---|---|
| 980 nm CW laser | ~10 mm | ~10 μm | Phosphorene QD solution | [38] |
| 980 nm CW laser | - | 8 μm | Antimonene solution | [83] |
| 974 nm CW laser | 1 cm | 13 μm | Bismuthene dispersion | [125] |
| 980 nm CW laser | - | 15 μm | Bismuthene dispersion | [124] |
| 1550 nm CW laser | 1 mm | 15 μm | Bismuthene dispersion | [121] |
| CW laser | 1 cm | 13 μm | Bismuthene dispersion | [115] |
| Red light | ~1 mm | ~16 μm | Bismuthene dispersion | [89] |

### 3.2    Balance twin-detector measurement

A balance twin-detector measurement setup was employed to measure nonlinear optical characterization of the SA as shown in Fig. 10(a). Typically, the twin detector setup consists of an ultrafast femtosecond laser source, a variable optical attenuator to control the input laser power, and a coupler dividing the section with the SA for the transmission-dependent and without the SA for the transmission-independent measurement [137][138]. The measurement was recorded with the optical power meters. Three important parameters could be retrieved from the transmittance or absorbance versus intensity curve as shown in Fig. 10(b), saturation intensity ($I_S$), modulation depth ($\Delta T$) and non-saturable loss ($\alpha_{ns}$). The $I_S$ could be measured when the transmittance of absorbance is half of the $\Delta T$, $\Delta T$ is the capability for saturable absorption that higher $\Delta T$ is typically preferable to generate shorter ultrashort pulses, and $\alpha_{ns}$ is the non-saturable loss, e.g. from the optical system or tapering loss that does not contributes to the saturable absorption process [139]. In some cases, average power (W) was utilized instead of intensity (MW/cm$^2$).

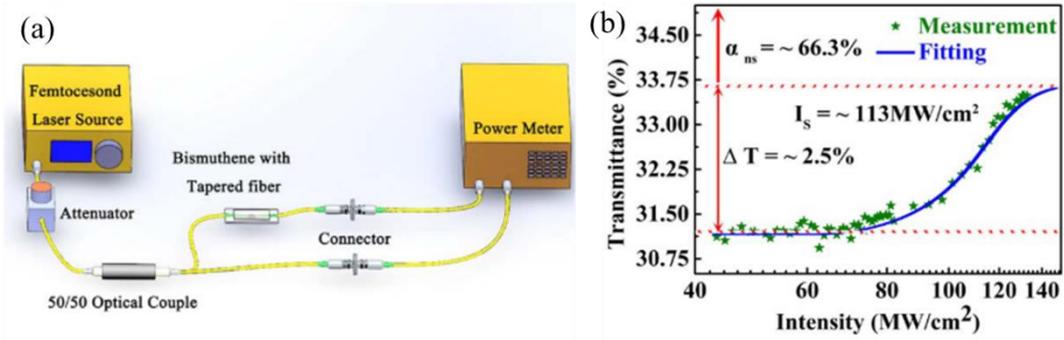

Fig. 10: (a) Twin-detector measurement setup and (b) Transmittance versus intensity curve, P. Guo et al. [121]. © IOP Publishing Ltd 2019.

The saturable absorption can be well fitted with a simplified two level model [10][59]. The absorption coefficient can be written as:

$$\alpha(N) = [\alpha_s / (1 + I/I_s)] + \alpha_{ns},$$

where α is the absorption coefficient, $\alpha_s$ is the saturable absorption components (ΔT), I is the photocarrier intensity, Is is the saturation intensity and $\alpha_{ns}$ is the non-saturable loss. The transmission coefficient is then calculated with 1 - α(N). The $I_S$, ΔT, and $\alpha_{ns}$ of the group-V mono-elemental SA are summarized in Table 2. Based on the study, bismuthene was more commonly used as the SA with the $I_S$ ranging from 0.3 MW/cm$^2$ to 113 MW/cm$^2$, whereas the ΔT was measured from 1% to 5.6%. In addition, further optimization for the bismuthene SA could be done to achieve $\alpha_{ns}$ lower than 50%.

Table 2: Summary of $I_S$, ΔT, and $\alpha_{ns}$ for group-V mono-elemental SA characterized through twin detector measurement setup.

| Material | Characterization Wavelength (nm) | $I_S$ / average power | ΔT (%) | $\alpha_{ns}$ (%) | Ref. |
|---|---|---|---|---|---|
| Phosphorene QD | 1031 | 1.35 MW/cm$^2$ | 8.1 | 2.5 | [2] |
| Phosphorene | 1566 | 3.5 mW | 13.3 | ~86.7 | [65] |
| Antimonene | 1568 | 10.8 mW | 6.4 | ~72.7 | [83] |
| Bismuthene | ~1550 | 0.4 MW/cm$^2$ | 3.8 | 78.7 | [125] |
| Bismuthene | ~1550 | 2.4 MW/cm$^2$ | 1 | ~78.8 | [115] |
| Bismuthene | ~1550 | ~30 MW/cm$^2$ | ~2.03 | ~82.5 | [110] |
| Bismuthene | ~1550 | ~110 MW/cm$^2$ | ~2.5 | ~66.5 | [122] |
| Bismuthene | 1550 | 48.2 MW/cm$^2$ | 5.6 | 62.3 | [89] |
| Bismuthene | ~1563 | 113 MW/cm$^2$ | ~2.5 | ~66.25 | [121] |
| Bismuthene | 1563.3 | 13 MW/cm$^2$ | 2.2 | ~50.7 | [124] |
| Bismuthene | 1563.3 | 0.3 MW/cm$^2$ | 2.4 | ~77.2 | [136] |

### 3.3 Z-scan measurement

The NLO response, nonlinear absorption coefficient, and refractive index of the 2D materials could also be characterized with an open aperture Z-scan technique with a femtosecond laser source as shown in Fig. 11(a) [50][82]. This setup was employed to measure the optical transmission of the sample as a function of intensity incident on the sample. The beam splitter (BS) acts as a free-space optical coupler that separates the attenuated laser power into two paths. In an open-aperture Z-scan setup, the aperture is a focusing lens that was placed before the sample to collect almost all of the transmitted light. This is beneficial to neglect small distortion of the light beam and the z-axis displacement dependent signal variation which is mainly caused by the nonlinear absorption. Contrarily, the close-aperture Z-scan setup only allows paraxial light to cross through thus the nonlinear refractive index of the sample could be measured [140].

An example of the ultrafast laser source used for the Z-scan setup is the mode-locked Ti:sapphire oscillator amplified with parametric amplifier at the seeded repetition rate of 1 kHz [50]. After the characterization from the Z-scan method, a phosphorene/graphene nano-heterostructure SA was measured with tunable ΔT from ~4.7 % to 9%, and tunable $I_S$ from 1.59 GW/cm$^2$ to 24.9 GW/cm$^2$ by increasing the phosphorene/graphene solution concentration [50]. In addition, another 800 nm, pulse width of 100 fs, and repetition rate of 1 kHz mode-locked Ti:sapphire oscillator-seed regenerative amplifier was employed for the Z-scan measurement of phosphorene SA, with the characterized result of ΔT at 14.2 %, $I_S$ at 774.4 GW/cm$^2$, and $α_{ns}$ of ~8% [59]. The same mode-locked Ti:sapphire oscillator with either 800 nm or 1500 nm wavelength was used to conduct the Z-scan measurement for an antimonene SA [82]. At the near-IR region, the nonlinear absorption coefficient of the antimonene was deduced as 7.88 ± 0.37 x 10$^{-15}$ cm/W at 1500 nm. In addition, the ΔT, $I_S$, and $α_{ns}$ for this antimonene at 1500 nm were measured as 12.8%, 63.7 GW/cm$^2$, and 2.03%, respectively as shown in Fig. 11(b) [82]. Another example of 1500 nm pulse laser source for Z-scan measurement is with 210 fs pulse duration [83]. The ΔT, $I_S$, and $α_{ns}$ for the antimonene SA were measured as 19.72 %, 15.10 GW/cm$^2$, and ~22.5% at 1500 nm, respectively. The pulse laser source for another Z-scan measurement is a 1064 nm, 10 Hz and 40 ps excitation laser for the characterization of an antimonene SA [102]. At this wavelength, the ΔT, $I_S$, and $α_{ns}$ of this antimonene SA is measured as 18.4%, 1.3 GW/cm$^2$, and 10%, respectively.

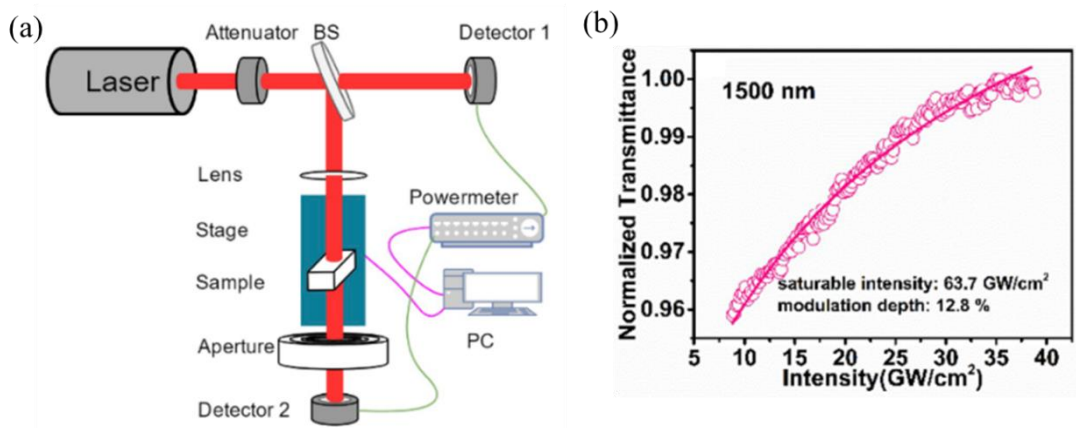

Fig. 11: (a) Schematic diagram for the Z-scan setup, F. Zhang et al. [82]. © Elsevier 2019.

## 4.0 Mode-locked fiber laser

The SA could be integrated into a fiber laser cavity to generate mode-locked pulses. A typical fiber laser cavity is constructed from either a ring or linear configuration. In this section, mode-locking in ytterbium-doped fiber laser (YDFL), erbium-doped fiber laser (EDFL), harmonic mode-locking and dual-wavelength mode-locking regimes will be discussed. Then, a critical review of group-V saturable absorber in mode-locked fiber laser will be tabulated in a summary table.

### 4.1 YDFL

A ring-cavity YDFL was constructed to achieve mode-locked laser as shown in Fig. 12(a). The laser cavity is composed of several components such as the pump laser with a wavelength of typically ~980 nm. Next, a wavelength division multiplexer (WDM) was employed for the multiplexing of pump wavelength (974 nm) and signal wavelength (1064 nm), a ytterbium (Yb)-doped fiber as the active gain medium, a polarization controller (PC) for the adjustment of the cavity birefringence in a non-polarization maintaining laser cavity, a SA (e.g. bismuthene) to initiate the mode-locking operation, a polarization-independent isolator (PI-ISO) to force unidirectional laser signal propagation, and finally a coupler with different coupling ratio, in which the lower portion of the optical signal was siphoned out for the measurement, whereas the larger portion of the optical signal reverted into the laser cavity to complete the setup. In typical, a mode-locked YDFL has net cavity group velocity dispersion (GVD) of normal dispersion with dissipative soliton characteristic which is mainly due to the positive dispersion of the optical fiber used in this wavelength region [115][124][125]. A simple declaration of dissipative soliton regime is the rectangular/triangular profile optical spectrum as shown in Fig. 12(b).

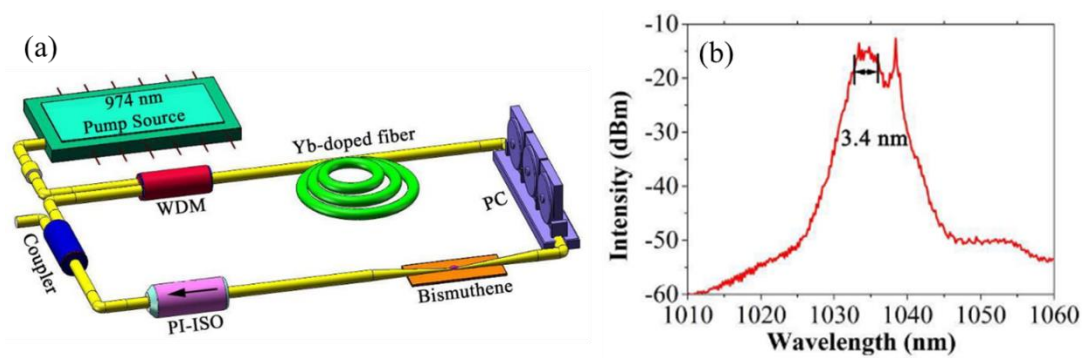

Fig. 12: (a) Schematic diagram of mode-locked fiber laser and (b) its optical spectrum, T. Feng et al. [125]. © Creative Commons Attribution 4.0 International License 2019.

### 4.2 EDFL

Apart from the YDFL, an EDFL has an emission wavelength of ~1550 nm due to the integration of erbium-doped fiber (EDF) as the active gain medium. The schematic diagram of a mode-locked EDFL is shown in Fig. 13(a) [83]. In most cases, the ring-cavity mode-locked EDFL integrated with the group-V mono-element SAs exhibits solitonic behaviour, e.g. with the presence of Kelly's sidebands in the optical spectrum as depicted in Fig. 13(b) [110]. The Kelly's sidebands denote the interplay between GVD and self-phase modulation (SPM) [141]. This is confirmed by the observation of narrow peaks superimposed on the soliton pulse spectrum arising from resonances between the soliton and dispersive wave components emitted after soliton perturbations [38]. The net cavity GVD is typically anomalous, such as ~-0.35 $ps^2$ [89] and -1.09 $ps^2$ [121][122]. Subsequently, two important parameters could be retrieved from the optical spectrum, which are the centre wavelength ($\lambda_c$) and the 3-dB spectral bandwidth ($\Delta\lambda$). These two parameters are useful for the calculation of time bandwidth produce (TBP), which is an indication of the ultrashort pulse performance especially on the chirping characteristic [142]. The TBP could be calculated based on the equation:

$$TBP = [(\Delta\lambda.c.\Delta\tau) / \lambda_c^2]$$

where c is the speed of light and $\Delta\tau$ is the pulse duration. The $\Delta\tau$ was measured typically in picosecond or femtosecond from an autocorrelator as shown in Fig. 13(c). Based on the TBP, the chirp-free values should be calculated as 0.441 and 0.315 for the two more commonly integrated fitted profiles, secant hyperbolic and Gaussian, respectively. The pulse is considered chirped when the measured TBP is larger than these values. Moreover, the pulse repetition rate is another important measurement from a passively mode-locked fiber laser, which is indicated by the period required by a laser cavity to complete a round-trip. The pulse repetition rate, typically in MHz range could be measured through an oscilloscope by converting the optical signal into electrical signal via a photodetector. An example of the pulse train with the fundamental repetition rate of 8.83 MHz is shown in Fig. 13(d). The stability of the pulse was then measured with the signal-to-noise ratio (*S/N*) via a radio frequency spectrum as shown in Fig. 13(e). The *S/N* was measured from the difference between the peak intensity and the ground intensity of the pulse at the fundamental repetition rate.

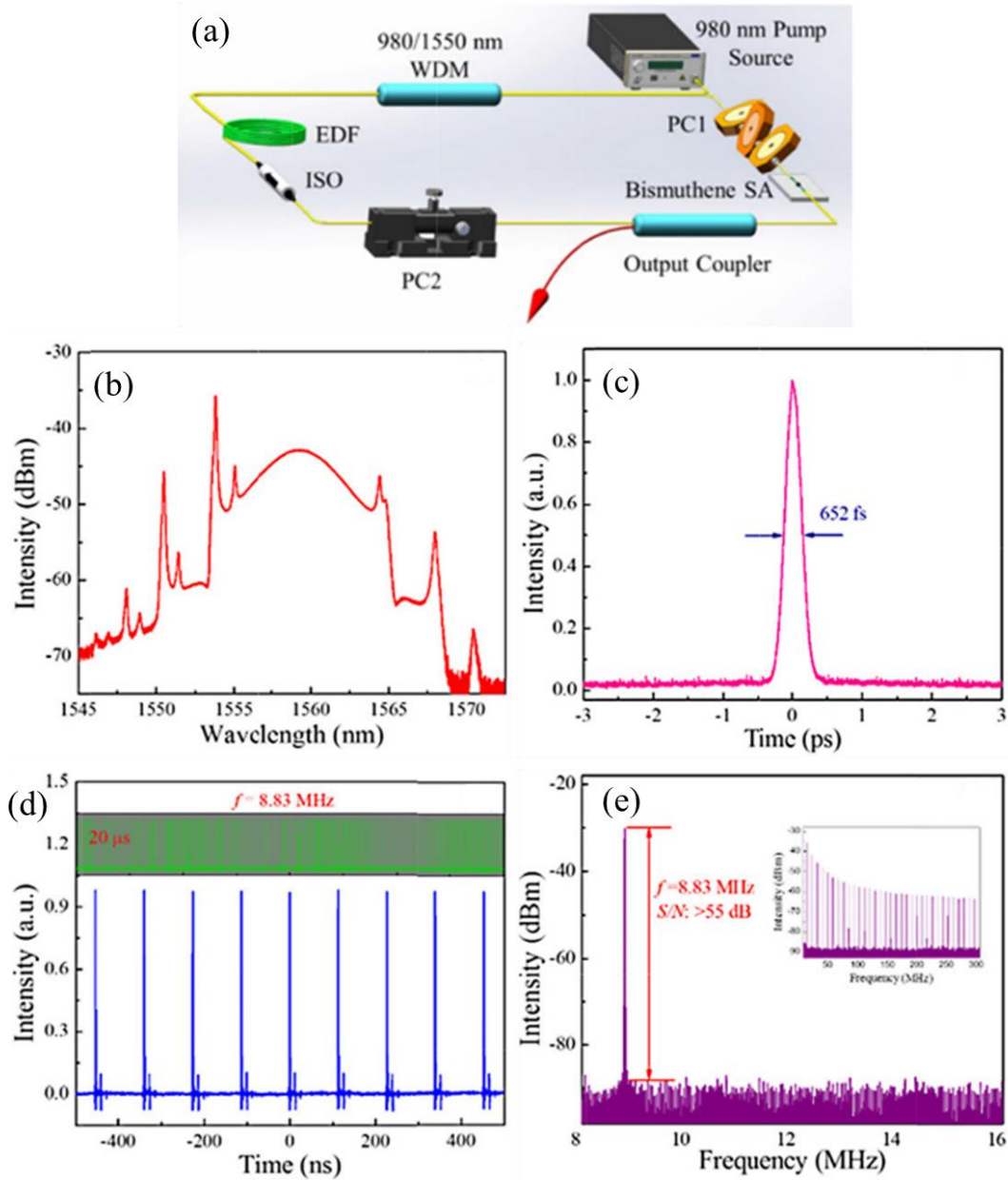

Fig. 13: (a) Schematic diagram for the mode-locked EDFL. (b) Optical spectrum, (c) Autocorrelation trace, (d) Pulse train; inset: Pulse train with 20 μs span, and (e) RF spectrum, L. Lu et al. [110]. © Wiley-VCH Verlag GmbH & Co. KGaA, Weinheim 2017.

## 4.3 Harmonic mode-locking

Apart from fundamental mode-locking, harmonic mode-locking allows multiple pulses to circulate in the laser cavity with equal temporal spacing. A harmonic mode-locking was achieved at 10$^{th}$ harmonic order, as well as dual-wavelength mode-locked operation at a centre wavelength of 1545.78 nm and 1555.3 nm with pump power of 81.85 mW [122]. When the pump power was further increased to 301.79 mW, the dual-wavelength mode-locked EDFL was observed at 1544.85 nm and 1554.53 nm, with the 3-dB spectral bandwidth of 1.9 nm and 1.82 nm, respectively [122]. The corresponding optical spectrum and pulse train for the harmonic mode-locking at 29.81, 81.85, and 301.7 mW are illustrated in Fig. 14(a)-(f). In addition, there are 4th and 165th harmonic of repetition rate observed at 8.49 MHz and 356 MHz for dual and single-wavelength ring cavity mode-locked EDFL employing antimonene SA, respectively [102]. It was reported that a phosphorene/graphene nanohetero-structure SA could generate shorter pulse duration of 148 fs with a lower solution concentration, rather than 820 fs with a higher solution concentration [50]. Nevertheless, the harmonic mode-locking with the highest harmonic order at 40th order could only be achieved with lower concentration of the phosphorene/graphene nanohetero-structure SA.

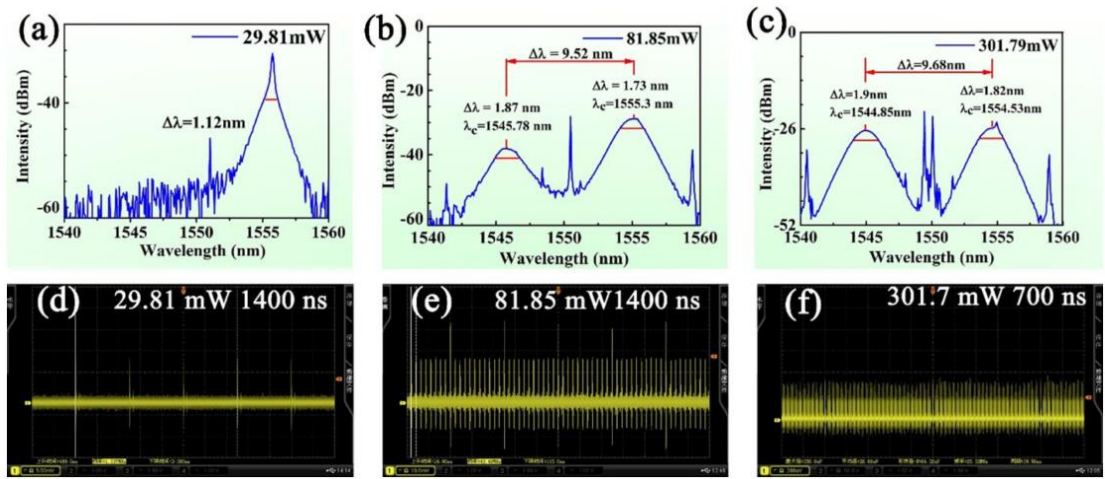

Fig. 14. Optical spectrum for (a) fundamental soliton, (b) dual-wavelength threshold, and (c) harmonic mode-locking. Pulse train for (d) fundamental soliton, (e) dual-wavelength threshold, and (f) harmonic mode-locking. P. Guo et al. [122]. © American Chemical Society 2020.

### 4.4    Summary

A summary of the mode-locked fiber laser employing group-V mono-element SA, including harmonic and single/dual-wavelength mode-locking operation regimes is summarized in this section. Table 3 summarizes the technical result in the recent five years by integrating the group-V SA into the mode-locked fiber laser with different fabrication techniques, materials, SA structure, deposition methods, and net cavity GVD. The $\lambda_c$, $\Delta\lambda$, $\Delta\tau$, TBP, and SNR are the abbreviation of centre wavelength, 3-dB spectral bandwidth, pulse duration, time bandwidth product, and signal-to-noise ratio, respectively.

Table 3: Summary of the overall performances for the mode-locked fiber laser with Group-V mono-elemental SAs.

| Year | Fabrication Method | Material | SA Structure | Deposition Technique | Net cavity GVD | $\lambda_c$ (nm) | $\Delta\lambda$ (nm) | $\Delta\tau$ | TBP | SNR (dB) | Repetition rate (MHz) | Ref. |
|---|---|---|---|---|---|---|---|---|---|---|---|---|
| 2017 | LPE | Phosphorene /graphene | D-shaped fiber | Immersion | Anomalous | 1531 | 19.4 | 148 fs | 0.35 (sech$^2$) | 58 | 1) 7.5 (1st order) 2) 295.6 (40th order) | [50] |
| 2017 | LPE | Phosphorene QD | Tapered fiber | Optical deposition | Anomalous | 1561.7 | 3 | 882 fs | 0.325 (sech$^2$) | 67 | 5.47 | [38] |
| 2017 | LPE | Antimonene | Tapered fiber | Optical deposition | Anomalous | 1557.68 | 4.84 | 552 fs | ~0.33 (sech$^2$) | 50 | 10.27 | [83] |
| 2017 | Sonochemical exfoliation | Bismuthene | Tapered fiber | Solution filling | Anomalous | 1559.18 | 4.64 | 652 fs | ~0.373 (sech$^2$) | 55 | 8.83 | [110] |
| 2018 | Sonochemical exfoliation | Bismuthene | Tapered fiber | Optical deposition | Normal | 1034.4 | 2.72 | 30.25 ns | 23.07 (Gauss.) | 45 | 21.74 | [124] |
| 2018 | Sonochemical exfoliation | Bismuthene | Tapered fiber | Optical deposition | Anomalous | 1561 | 14.4 | 193 fs | ~0.342 (sech$^2$) | 55 | 8.85 | [89] |
| 2019 | LPE | Antimonene | Fiber ferrule | Dipping and drying | Anomalous | 1) 1564 (single) 2) 1564 (165th harmonics) | 1) 0.28 (single) 2) 0.278 (165th harmonics) | 1) ~1.73 ns (single) 2) ~953 ps (165th harmonics) | 1) 59.41 (Gauss.) 2) 32.49 (Gauss.) | - | 1) 2.16 (single) 2) 356 (165th harmonics) | [102] |
| 2019 | Sonochemical exfoliation | Bismuthene | Tapered fiber | Drop casting | Anomalous | 1557.5 | 10.35 | 621.5 fs | ~0.796 (sech$^2$) | 25 | 22.74 | [136] |
| 2019 | Sonochemical exfoliation | Bismuthene | Tapered fiber | Optical deposition | Anomalous | 1531 | 1.6 | 1.303 ps | ~1.67 (sech$^2$) | 56.54 | 4 | [121] |
| 2019 | Sonochemical exfoliation | Bismuthene | Tapered fiber | Laser injection | Normal | 1035.8 | 3.4 | 1) 1.55 ps (Coherent Peak) 2) 54.19 ps (Envelope) | 1) 1.47 (Gauss.) 2) 51.52 (Gauss.) | 29.6 | 21.74 | [125] |
| 2020 | Bath sonication and probe sonication | Bismuthene | Tapered fiber | Solution filling | Anomalous | 1) 1544.85 (short $\lambda_c$) 2) 1554.53 (Long $\lambda_c$) | 1) 1.9 (short $\lambda_c$) 2) 1.82 (Long $\lambda_c$) | 1) 1.3 ps (short $\lambda_c$) 2) 1.3 ps (Long $\lambda_c$) | 1) ~0.31 (sech$^2$) 2) ~0.29 (sech$^2$) | < 30 | 1) 4 (short $\lambda_c$) 2) 4 (Long $\lambda_c$) (40 MHz at 10th harmonics) | [122] |
| 2020 | Dispersion | Bismuthene | Tapered fiber | Optical deposition | Normal | 1035.8 | 3.4 | 1) 1.55 ps (Coherent peak) 2) 76.62 ps (Envelope) | 1) 1.47 (Gauss.) 2) 72.84 (Gauss.) | 29.6 | 21.74 | [115] |

## 5.0 Challenges and Recommendations

The challenges of the phosphorene, arsenene, antimonene, and bismuthene will be discussed individually. The main challenge of the phosphorene is its oxidation issue [143]. Therefore, the phosphorene is difficult to maintain its initial characteristics due to the exposure to the atmospheric environment, even within 30 minutes of preparation [144]. This could influence the quality of SA and thus the stability of the mode-locked fiber laser. There are several approaches to prevent or reduce the rate of oxidation for phosphorene. An example is the encapsulation of phosphorene with $AlO_x$ through atomic layer deposition which was proven with stability of more than two weeks [145]. In addition, the encapsulation of phosphorene layer between boron or nitrogen (BN) layers to become hetero-structure was proven with preservation of at least two months after the fabrication [146]. Moreover, the dispersion of phosphorene QD into NMP was proven with excellent environmental stability for 6 months without degradation [38]. In contrary to the challenges, the phosphorene was validated with broadband absorption, especially an absorption band at ~1900 nm according to Fig. 10(f). This could be an opportunity to investigate the application of phosphorene SA to achieve thulium-doped mode-locked fiber laser, which is very useful for the eye-safety application due to the low water absorption peak at this wavelength region [147]. Meanwhile, both arsenene and antimonene face the same challenge, which are their bandgap values is still in the prediction stage. Therefore, instead of predicting the bandgap according to density-function-theory calculation, experimental work on validating these bandgap values would fulfil this research gap perfectly. In fact, both arsenene and antimonene shows bandgap tunability advantage from metal to semiconducting by reducing the material thickness. For instance, the zero bandgap properties were exhibited by trilayer arsenene and bilayer antimonene according to Fig. 3(a) and Fig. 7(b), respectively. The zero bandgap is a remarkable advantage, likewise to graphene with ultra-broadband wavelength ranging up to terahertz operation [148][149]. Nevertheless, very little research was done on mode-locked fiber laser with antimonene SA, and surprisingly, none of the mode-locked fiber laser was ever reported with arsenene SA. The authors believe the zero bandgap properties of the trilayer arsenene and bilayer antimonene should attract some research interest in the future for their potential as the SA to generate mode-locked fiber laser in near-IR region and other wavelength region. In addition, both arsenene and antimonene has higher resistance to oxidation than phosphorene. Bismuthene is a maturely studied material that has been demonstrated with quite a lot of mode-locked fiber laser researches in near IR-region of ~1064 nm and ~1550 nm wavelength according to Table 3. In conjunction to arsenene and antimonene, bismuthene has a closely zero bandgap when it is at least in the bilayer form as presented in Fig. 7(b). This contributes to the feasibility of bismuthene as a SA in ~2000 nm wavelength. Based on the Table 3, the research is still pending with a first demonstration of bismuthene SA in thulium- or holmium-doped mode-locked fiber laser. Apart from that, the demonstration of the classic mode-locked fiber laser could facilitate the investigation of several other regimes, such as soliton rains, vector soliton, high energy rectangular pulses based on dissipative soliton resonance, optical rogue waves and so on by the design of the fiber laser cavity. The real-time observation of the mode-locked fiber laser phenomena could also be realized with its frequency domain measurement via time-stretched dispersive Fourier transform [150][151] and temporal domain measurement via time lens technique [152][153]. In addition, the monolayer bismuthene with 0.3 to 0.5 eV bandgap was proven to be topologically non-trivial, thus rendering its significant values as 2D topological insulator. Subsequently, topological insulator has been readily and extensively demonstrated as the SA in the previous mode-locked fiber laser research since one of its earliest demonstration in 2015 [154].

## 6.0   Conclusion and outlook

In summary, we had successfully reviewed the recent development of group-V mono-element 2D material as saturable absorber for near-IR mode-locked fiber laser. Four group-V elements in its 2D form, phosphorene, arsenene, antimonene and bismuthene are studied in terms of its material properties such as band structure, atomic structure, lattice constant, inter-layer distance and so on. The most stable regime of the group-V monolayer elements is puckered (α-phase) for phosphorene and buckled (β-phase) for the arsenene, antimonene and bismuthene. In addition, the synthesis method of these 2D materials had been reviewed with thorough procedures of several approaches such as ME, LPE, sonochemical exfoliation, bath sonication and probe sonication, etc. In most approaches, centrifugation appears to be the last step before the final products were synthesized. Next, the synthesized material was characterized in terms of its bandgap energy with photoluminescence spectrum, Raman characteristic peaks with Raman spectrum, crystallinity peaks with XPS spectrum, absorption spectrum with UV-VIS-NIR spectrum, nanosheet structure with SEM image, lattice plane and interplanar distance with TEM and HRTEM, and the lateral height with AFM image. These materials are then employed for the SA fabrication through several platforms, such as fiber ferrule, D-shaped fiber and tapered fiber. Among these platforms, tapered fiber is most commonly implemented due to its advantages such as high optical damage threshold, excellent heat dissipation mechanism, large evanescent field for NLO interaction with the SA materials, insensitive to polarization-dependent loss, and fabrication maturity via flame-brush technique. The NLO parameters, particularly $I_s$, $\Delta T$, and $\alpha_{ns}$ of the SAs were characterized with twin-detector method and Z-scan method. Based on Table 2, the high $\alpha_{ns}$ of the proposed SA could be further improved and optimized in the future. Next, these SAs were integrated into near-IR fiber laser cavity to achieve mode-locked operation. According to Table 3, the demonstration of arsenene SA in near-IR region, ~1900 nm to 2000 nm mode-locked fiber laser with all group-V SAs and low SNR of ~30 dB in most recent bismuthene-SA mode-locked fiber laser should be given a consideration to fulfil these research gaps. Finally, the challenges of this research were summarized and resolved with some recommendations. The study of this review hopefully could provide a better insight to the readers regarding the implementation of group-V 2D mono-element materials as the SA in near-IR mode-locked fiber laser.

## Acknowledgement

This work was funded by the International Postdoctoral Exchange Fellowship Program (Talent-Introduction Program), Office of China Postdoc Council (OCPC).

## References

[1] A. J. DeMaria, D. A. Stetser, and H. Heynau, "Self mode-locking of lasers with saturable absorber," *Appl. Phys. Lett.*, vol. 8, no. 174, pp. 1964–1967, 1966.

[2] B. Zhang, F. Lou, R. Zhao, J. He, J. Li, X. Su, J. Ning, and K. Yang, "Exfoliated layers of black phosphorus as saturable absorber for ultrafast solid-state laser," *Opt. Lett.*, vol. 40, no. 16, pp. 3691–3694, 2015.

[3] U. Keller, "Recent developments in compact ultrafast lasers," *Nature*, vol. 424, pp. 831–838, 2003.

[4] M. E. Fermann and I. Hartl, "Ultrafast fibre lasers," *Nat. Photonics*, vol. 7, pp. 868–874, 2013.

[5] J. Bogusławski, Y. Wang, H. Xue, X. Yang, D. Mao, X. Gan, X. Ren, J. Zhao, Q. Dai, G. Sobon, J. Sotor, and Z. Sun, "Graphene Actively Mode-Locked Lasers," *Adv. Funct. Mater.*, vol. 28, p. 1801539, 2018.

[6] L. G. Wright, D. N. Christodoulides, and F. W. Wise, "Spatiotemporal mode-locking in multimode fiber lasers," *Science*, vol. 358, pp. 94–97, 2017.

[7] Z. Sun, D. Popa, T. Hasan, F. Torrisi, F. Wang, E. J. R. Kelleher, J. C. Travers, V. Nicolosi, and A. C. Ferrari, "A stable, wideband tunable, near transform-limited, graphene-mode-locked, ultrafast laser," *Nano Res.*, vol. 3, no. 9, pp. 653–660, 2010.

[8] G. Wang, A. A. Baker-Murray, and W. J. Blau, "Saturable Absorption in 2D Nanomaterials and Related Photonic Devices," *Laser Photon. Rev.*, vol. 13, p. 1800282, 2019.

[9] F. Meng, M. D. Thomson, F. Bianco, A. Rossi, D. Convertino, A. Tredicucci, C. Coletti, and H. G. Roskos, "Saturable absorption of femtosecond optical pulses in multilayer turbostratic graphene," *Opt. Express*, vol. 24, no. 14, pp. 15261–15273, 2016.

[10] Q. Bao, H. Zhang, Y. Wang, Z. Ni, Y. Yan, Z. X. Shen, K. P. Loh, and D. Y. Tang, "Atomic-layer graphene as a saturable absorber for ultrafast pulsed lasers," *Adv. Funct. Mater.*, vol. 19, no. 19, pp. 3077–3083, 2009.

[11] F. Bonaccorso, Z. Sun, T. Hasan, and A. C. Ferrari, "Graphene photonics and optoelectronics," *Nat. Photonics*, vol. 4, pp. 611–622, 2010.

[12] L. Miao, Y. Jiang, S. Lu, B. Shi, C. Zhao, H. Zhang, and S. Wen, "Broadband ultrafast nonlinear optical response of few-layers graphene: toward the mid-infrared regime," *Photon. Res.*, vol. 3, no. 5, pp. 214–219, 2015.

[13] S. Das Sarma, S. Adam, E. H. Hwang, and E. Rossi, "Electronic transport in

two-dimensional graphene," *Rev. Mod. Phys.*, vol. 83, no. 2, pp. 408–470, 2011.

[14] A. Martinez and Z. Sun, "Nanotube and graphene saturable absorbers for fibre lasers," *Nat. Photonics*, vol. 7, pp. 842–845, 2013.

[15] X. Lin, J. C. Lu, Y. Shao, Y. Y. Zhang, X. Wu, J. B. Pan, L. Gao, S. Y. Zhu, K. Qian, Y. F. Zhang, D. L. Bao, L. F. Li, Y. Q. Wang, Z. L. Liu, J. T. Sun, T. Lei, C. Liu, J. O. Wang, K. Ibrahim, D. N. leonard, W. Zhou, H. M. Guo, Y. L. Wang, S. X. Du, S. T. Pantelides, and H. -J. Gao, "Intrinsically patterned two-dimensional materials for selective adsorption of molecules and nanoclusters," *Nat. Mater.*, vol. 16, pp. 717–722, 2017.

[16] X. Niu, Y. Yi, L. Meng, H. Shu, and Y. Pu, "Two-Dimensional Phosphorene, Arsenene, Antimonene Quantum Dots: Anomalous Size-Dependent Behaviors of Optical Properties," *J. Phys. Chem. C*, vol. 123, no. 42, pp. 25775–25780, 2019.

[17] Y. Zhang, T. -R. Chang, B. ZHou, Y. -T. Cui, H. Yan, Z. Liu, F. Schmitt, J. lee, R. Moore, Y. Chen, H. Lin, H. -T. Jeng, S. -K. Mo, Z. Hussain, A. Bansil, and Z. -X. Shen, "Direct observation of the transition from indirect to direct bandgap in atomically thin epitaxial $MoSe_2$," *Nat. Nanotechnol.*, vol. 9, pp. 111–115, 2014.

[18] Z. Wu and J. Hao, "Electrical transport properties in group-V elemental ultrathin 2D layers," *npj 2D Mater. Appl.*, vol. 4, no. 4, pp. 1–13, 2020.

[19] S. Zhang, S. Guo, Z. Chen, and Y. Wang, "Recent progress in 2D group-VA semiconductors: from theory to experiment," *Chem. Rev.*, vol. 47, no. 3, pp. 982–1021, 2018.

[20] G. Qin, and Z. Qin, "Negative Poisson' s ratio in two-dimensional honeycomb structures," *npj Comput. Mater.*, vol. 51, pp. 1–6, 2020.

[21] S. Zhang, Z. Yan, Y. Li, Z. Chen, and H. Zeng, "Atomically Thin Arsenene and Antimonene: Semimetal-Semiconductor and Indirect-Direct Band-Gap Transitions," *Angew. Chemie*, vol. 127, no. 10, pp. 3155–3158, 2015.

[22] S. Fukuoka, T. Taen, and T. Osada, "Electronic Structure and the Properties of Phosphorene and Few-Layer Black Phosphorus," *J. Phys. Soc. Japan*, vol. 84, no. 12, p. 121004, 2015.

[23] J. Qiao, X. Kong, Z. Hu, F. Yang, and W. Ji, "High-mobility transport anisotropy and linear dichroism in few-layer black phosphorus," *Nat. Commun.*, vol. 5, p. 4475, 2014.

[24] J. Xie, M. S. Si, D. Z. Yang, Z. Y. Zhang, and D. S. Xue, "A theoretical study of blue phosphorene nanoribbons based on first-principles calculations A theoretical study of blue phosphorene nanoribbons based on first-principles calculations," *J. Appl. Phys.*, vol. 116, no. 7, p. 073704, 2014.

[25] M. Wada, S. Murakami, F. Freimuth, and G. Bihlmayer, "Localized edge states in two-dimensional topological insulators : Ultrathin Bi films," *Phys. Rev. B*,

vol. 83, no. 12, p. 121310, 2011.

[26] S. Murakami, "Quantum Spin Hall Effect and Enhanced Magnetic Response by Spin-Orbit Coupling," *Phys. Rev. Lett.*, vol. 97, no. 23, p. 236805, 2006.

[27] S. Yamashita, "Nonlinear optics in carbon nanotube, graphene, and related 2D materials Nonlinear optics in carbon nanotube, graphene, and related 2D materials," *APL Photonics*, vol. 4, p. 034301, 2019.

[28] P. C. Debnath and D. Il Yeom, "Ultrafast fiber lasers with low-dimensional saturable absorbers: Status and prospects," *Sensors*, vol. 21, p. 3676, 2021.

[29] J. W. You, S. R. Bongu, Q. Bao, and N. C. Panoiu, "Nonlinear optical properties and applications of 2D materials: Theoretical and experimental aspects," *Nanophotonics*, vol. 8, no. 1, pp. 63–97, 2018.

[30] P. Vishnoi, K. Pramoda, and C. N. R. Rao, "2D Elemental Nanomaterials Beyond Graphene," *ChemNanoMat*, vol. 5, pp. 1–31, 2019.

[31] X. Liu, Q. Gao, Y. Zheng, D. Mao, and J. Zhao, "Recent progress of pulsed fiber lasers based on transition-metal dichalcogenides and black phosphorus saturable absorbers," *Nanophotonics*, vol. 9, no. 8, pp. 2215–2231, 2020.

[32] M. Zhang, Q. Wu, F. Zhang, L. Chen, X. Jin, and Y. Hu, "2D Black Phosphorus Saturable Absorbers for Ultrafast Photonics," *Adv. Opt. Mater.*, vol. 7, p. 1800224, 2018.

[33] L. Zhang, B. Wang, Y. Zhou, C. Wang, X. Chen, and H. Zhang, "Synthesis Techniques, Optoelectronic Properties, and Broadband Photodetection of Thin-Film Black Phosphorus," *Adv. Opt. Mater.*, vol. 8, no. 15, p. 2000045, 2020.

[34] C. Ma, W. Huang, Y. Wang, J. Adams, Z. Wang, J. Liu, Y. Song, Y. Ge, Z. Guo, L. Hu, and H. Zhang, "MXene saturable absorber enabled hybrid mode-locking technology: A new routine of advancing femtosecond fiber lasers performance," *Nanophotonics*, vol. 9, no. 8, pp. 2451–2458, 2020.

[35] C. Ma, C. Wang, B. Gao, J. Adams, and H. Zhang, "Recent progress in ultrafast lasers based on 2D materials as a saturable absorber," *Appl. Phys. Rev.*, vol. 6, no. 4, p. 041304, 2019.

[36] G. Liu, J. Yuan, Y. Lyu, T. Wu, Z. Li, M. Zhu, F. Zhang, F. Zing, W. Zhang, H. Zhang, and S. Fu, "Few-layer silicene nanosheets as saturable absorber for subpicosecond pulse generation in all-fiber laser," *Opt. Laser Technol.*, vol. 131, p. 106397, 2020.

[37] H. Mu, Y. Liu, S. R. Bongu, X. Bao, L. Li, S. Xiao, J. Zhuang, C. Liu, Y. Huang, Y. Dong, K. Helmerson, K. Wang, G. Liu, Y. Du, and Q. Bao, "Germanium Nanosheets with Dirac Characteristics as a Saturable Absorber for Ultrafast Pulse Generation," *Adv. Mater.*, vol. 33, p. 2101042, 2021.

[38] J. Du, M. Zhang, Z. Guo, J. Chen, X. Zhu, G. Hu, P. Peng, Z. Zheng, and H. ZHang, "Phosphorene quantum dot saturable absorbers for ultrafast fiber lasers," *Sci. Rep.*, vol. 7, p. 42357, 2017.

[39] S. Zhang, S. Guo, Z. Chen, Y. Wang, H. Gao, J. Gómez-Herrero, P. Ares, F. Zamora, Z. Zhu, and H. Zeng, "Recent progress in 2D group-VA semiconductors: From theory to experiment," *Chem. Soc. Rev.*, vol. 47, no. 3, pp. 982–1021, 2018.

[40] J. Dai and X. C. Zeng, "Structure and stability of two dimensional phosphorene with O or NH functionalization," *RSC Adv.*, vol. 4, no. 89, pp. 48017–48021, 2014.

[41] M. Y. Yao, F. Zhu, C. Q. Han, D. D. Guan, C. Liu, D. Qian, and J. F. Jia, "Topologically nontrivial bismuth(111) thin films," *Sci. Rep.*, vol. 6, p. 21326, 2016.

[42] H. Thurn and H. Kerbs, "Crystal structure of violet phosphorus.," *Angew. Chemie Int. Ed.*, vol. 5, no. 12, pp. 1047–1048, 1966.

[43] S. Appalakondaiah, V. Vaitheeswaran, S. Lebègue, N. E. Christensen, and A. Svane, "Effect of van der Waals interactions on the structural and elastic properties of black phosphorus," *Phys. Rev. B*, vol. 86, no. 3, p. 035105, 2012.

[44] S. Wu, K. S. Hui, and K. N. Hui, "2D Black Phosphorus: from Preparation to Applications for Electrochemical Energy Storage," *Adv. Sci.*, vol. 5, p. 1700491, 2018.

[45] L. Liang, J. Wang, W. Lin, B. G. Sumpter, V. Meunier, and M. Pan, "Electronic Bandgap and Edge Reconstruction in Phosphorene Materials," *Nano Lett.*, vol. 14, no. 11, pp. 6400–6406, 2014.

[46] D. Li, H. Jussila, L. Karvonen, G. Ye, H. Lipsanen, X. Chen, and Z. Sun, "Polarization and Thickness Dependent Absorption Properties of Black Phosphorus : New Saturable Absorber for Ultrafast Pulse Generation," *Sci. Rep.*, vol. 5, p. 15899, 2015.

[47] A. Castellanos-gomez, L. Vivarelli, E. Prada, J. O Island, K. L. Narasimha-Acharya, S. I. Blanter, D. J. Groenendijk, M. Buscema, G. A. Steele, J. V. Alvarez, H. W. Zandbergen, J. J. Palacios, and H. S. J. Van Der Zant, "Isolation and characterization of few-layer black phosphorus," *2D Mater.*, vol. 1, p. 025001, 2014.

[48] J. Zhang, X. Yu, W. Han, B. Lv, X. Li, S. Xiao, Y. Gao, and J. He, "Broadband spatial self-phase modulation of black phosphorous," *Opt. Lett.*, vol. 41, no. 8, pp. 1704–1707, 2016.

[49] J. O. Island, G. A. Steele, H. S. J. Van Der Zant, and A. Castellanos-gomez, "Environmental instability of few-layer black phosphorus," *2D Mater.*, vol. 2, p. 011002, 2015.

[50] S. Liu, Z. Li, Y. Ge, H. Wang, R. Yue, X. Jiang, J. Li, Q. Wen, and H. Zhang, "Graphene/phosphorene nano-heterojunction: facile synthesis, nonlinear optics, and ultrafast photonics applications with enhanced performance," *Photonics Res.*, vol. 5, no. 6, pp. 662-668, 2017.

[51] L. Li, Y. Yu, G. J. Ye, Q. Ge, X. Ou, H. Wu, D. Feng, X. H. Chen, and Y.

Zhang, "Black phosphorus field-effect transistors," *Nat. Nanotechnol.*, vol. 9, pp. 372–377, 2014.

[52] T. Hu, Y. Han, and J. Dong, "Mechanical and electronic properties of monolayer and bilayer phosphorene under uniaxial and isotropic strains," *Nanotechnology*, vol. 25, p. 455703, 2014.

[53] L. Kou, C. Chen, and S. C. Smith, "Phosphorene: Fabrication, Properties and Applications," *J. Phys. Chem. Lett.*, vol. 6, no. 14, pp. 2794–2805, 2015.

[54] H. Liu, A. T. Neal, Z. Zhu, Z. Luo, X. Xu, D. Tománek, and P. D. Ye, "Phosphorene: An Unexplored 2D Semiconductor with a High Hole Mobility," *ACS Nano*, vol. 8, no. 4, pp. 4033–4041, 2014.

[55] B. Li, C. Lai, G. Zeng, D. Huang, L. Qin, and M. Zhang, "Black Phosphorus, a Rising Star 2D Nanomaterial in the Post-Graphene Era: Synthesis, Properties, Modifications, and Photocatalysis Applications," *Small*, vol. 15, p. 1804565, 2019.

[56] S. B. Lu, L. L. Miao, Z. N. Guo, X. Qi, C. J. Zhao, H. Zhang, S. C. Wen, D. Y. Tang, and D. Y. Fan, "Broadband nonlinear optical response in multi-layer black phosphorus: an emerging infrared and mid-infrared optical material," *Opt. Express*, vol. 23, no. 9, pp. 3370–3377, 2015.

[57] Y. Cai, G. Zhang, and Y. W. Zhang, "Layer-dependent band alignment and work function of few-layer phosphorene," *Sci. Rep.*, vol. 4, p. 6677, 2014.

[58] V. Paolucci, G. D'Olimpio, L. Lozzi, A. M. Mio, L. Ottaviano, M. Nardone, G. Nicotra, P. L. Cornec, C. Cantalini, and A. Politano, "Sustainable Liquid-Phase Exfoliation of Layered Materials with Nontoxic Polarclean Solvent," *ACS Sustain. Chem. Eng.*, vol. 8, no. 51, pp. 18830–18840, 2020.

[59] Z. Guo, H. Zhang, S. Lu, Z. Wang, S. Tang, J. Shao, Z. Sun, H. Xie, H. Wang, X. F. Yu, P. K. Chu, "From Black Phosphorus to Phosphorene : Basic Solvent Exfoliation, Evolution of Raman Scattering, and Applications to Ultrafast Photonics," *Adv. Funct. Mater.*, vol. 25, pp. 6996–7002, 2015.

[60] J. R. Brent, N. Savjani, E. A. Lewis, S. J. Haigh, D. J. Lewis, and P. O'Brien, "Production of few-layer phosphorene by liquid exfoliation of black phosphorus," *Chem. Commun.*, vol. 50, no. 87, pp. 13338–13341, 2014.

[61] P. Yasaei, B. Kumar, T. Foroozan, C. Wang, M. Asadi, D. Tuschel, J. E. Indacochea, R. F. Klie, and A. S. Khojin, "High-Quality Black Phosphorus Atomic Layers by Liquid-Phase Exfoliation," *Adv. Mater.*, vol. 27, no. 11, pp. 1887–1892, 2015.

[62] D. Hanlon, C. Backes, E. Doherty, C. S. Cucinotta, N. C. Berner, C. Boland, K. Lee, A. Harvey, P. Lynch, Z. Gholamvand, S. Zhang, K. Wang, G. Moynihan, A. Pokle, Q. M. Ramasse, N. McEvoy, W. J. Blau, J. Wang, G. Abellan, F. Hauke, A. Hirsch, S. Sanvito, D. D. O'Regan, G. S. Duesberg, V. Nicolosi, and J. N. Coleman, "Liquid exfoliation of solvent-stabilized few-layer black phosphorus for applications beyond electronics," *Nat. Commun.*, vol. 6, p. 8563, 2015.

[63] A. H. Woomer, T. W. Farnsworth, J. Hu, R. A. Wells, C. L. Donley, and S. C. Warren, "Phosphorene: Synthesis, Scale-up, and Quantitative Optical Spectroscopy," *ACS Nano*, vol. 9, no. 9, pp. 8869–8884, 2015.

[64] W. A. Crichton, M. Mezouar, G. Monaco, and S. Falconi, "Phosphorus: New in situ powder data from large-volume apparatus," *Powder Diffr.*, vol. 18, no. 2, pp. 155–158, 2003.

[65] J. Zheng, X. Tang, Z. Yang, Z. Liang, Y. Chen, and K. Wang, "Few-Layer Phosphorene-Decorated Microfiber for All-Optical Thresholding and Optical Modulation," *Adv. Opt. Mater.*, vol. 5, no. 9, p. 1700026, 2017.

[66] A. R. Baboukani, I. Khakpour, V. Drozd, A. Allagui, and C. Wang, "Single-step exfoliation of black phosphorus and deposition of phosphorene via bipolar electrochemistry for capacitive energy storage application," *J. Mater. Chem. A*, vol. 7, pp. 25548–25556, 2019.

[67] A. Ambrosi and M. Pumera, "Exfoliation of layered materials using electrochemistry," *Chem. Soc. Rev.*, vol. 47, p. 7213, 2018.

[68] Y. Yang, H. Hou, G. Zou, W. Shi, H. Shuai, J. Li, and X. Ji, "Electrochemical exfoliation of graphene-like two-dimensional nanomaterials," *Nanoscale*, vol. 11, pp. 16–33, 2019.

[69] A. Ambrosi, Z. Sofer, and M. Pumera, "Electrochemical Exfoliation of Layered Black Phosphorus into Phosphorene," *Angew. Chemie*, vol. 129, no. 35, pp. 10579–10581, 2017.

[70] S. Mardanya, V. K. Thakur, S. Bhowmick, and A. Agarwal, "Four allotropes of semiconducting layered arsenic that switch into a topological insulator via an electric field: Computational study," *Phys. Rev. B*, vol. 94, p. 035423, 2016.

[71] C. Kamal and M. Ezawa, "Arsenene: Two-dimensional buckled and puckered honeycomb arsenic systems," *Phys. Rev. B*, vol. 91, p. 085423, 2015.

[72] Y. Wang, C. Zhang, W. Ji, R. Zhang, and P. Li, "Tunable quantum spin Hall effect via strain in two-dimensional arsenene monolayer," *J. Phys. D. Appl. Phys.*, vol. 49, no. 5, p. 055305, 2016.

[73] J. Shah, W. Wang, H. M. Sohail, and R. I. G. Uhrberg, "Experimental evidence of monolayer arsenene: An exotic two-dimensional semiconducting material," *2D Mater.*, vol. 7, p. 025013, 2020.

[74] H. Tsai, S. W. Wang, C. H. Hsiao, C. W. Chen, H. Ouyang, Y. L. Chueh, H. C. Kuo, and J. H. Liang, "Direct Synthesis and Practical Bandgap Estimation of Multilayer Arsenene Nanoribbons," *Chem. Mater.*, vol. 28, no. 2, pp. 425–429, 2016.

[75] R. Gusmão, Z. Sofer, D. Bouša, and M. Pumera, "Pnictogen (As, Sb, Bi) Nanosheets for Electrochemical Applications Are Produced by Shear Exfoliation Using Kitchen Blenders," *Angew. Chemie*, vol. 129, no. 46, pp. 14609–14614, 2017.

[76] R. Gui, H. Jin, Y. Sun, X. Jiang, and Z. Sun, "Two-dimensional group-VA

nanomaterials beyond black phosphorus: Synthetic methods, properties, functional nanostructures and applications," *J. Mater. Chem. A*, vol. 7, no. 45, pp. 25712–25771, 2019.

[77] P. Vishnoi, M. Mazumder, S. K. Pati, and C. N. R. Rao, "Arsenene nanosheets and nanodots," *New J. Chem.*, vol. 42, no. 17, pp. 14091–14095, 2018.

[78] N. Antonatos, V. Mazánek, P. Lazar, J. Sturala, and Z. Sofer, "Acetonitrile-assisted exfoliation of layered grey and black arsenic: contrasting properties," *Nanoscale Adv.*, vol. 2, pp. 1282–1289, 2020.

[79] X. Wang, Y. Hu, J. Mo, J. Zhang, Z. Wang, W. Wei, H. Li, Y. Xu, J. Ma, J. Zhao, Z. Jin, and Z. Guo, "Arsenene: A Potential Therapeutic Agent for Acute Promyelocytic Leukaemia Cells by Acting on Nuclear Proteins," *Angew. Chemie - Int. Ed.*, vol. 59, no. 13, pp. 5151–5158, 2020.

[80] J. Sturala, Z. Sofer, and M. Pumera, "Coordination chemistry of 2D and layered gray arsenic: photochemical functionalization with chromium hexacarbonyl," *NPG Asia Mater.*, vol. 11, no. 42, pp. 1–7, 2019.

[81] P. Ares, F. A. Galindo, D. R. S. Miguel, D. A. Aldave, S. D. Tendero, M. Alcamí, F. Martin, J. G. Herrero, and F. Zamora, "Mechanical Isolation of Highly Stable Antimonene under Ambient Conditions," *Adv. Mater.*, vol. 28, no. 30, pp. 6332–6336, 2016.

[82] F. Zhang, X. Jiang, Z. He, W. Liang, S. Zu, and H. Zhang, "Third-order nonlinear optical responses and carrier dynamics in antimonene," *Opt. Mater.*, vol. 95, p. 109209, 2019.

[83] Y. Song, Z. Liang, X. Jiang, Y. Chen, Z. Li, L. Lu, Y. Ge, K. Wang, J. Zheng, S. Lu, J. Ji, and H. Zhang, "Few-layer antimonene decorated microfiber: ultra-short pulse generation and all-optical thresholding with enhanced long term stability," *2D Mater.*, vol. 4, no. 4, p. 045010, 2017.

[84] Y. Shao, Z. L. Liu, C. Cheng, X. Wu, H. Liu, C. Liu, J. Wang, S. Y. Zhu, Y. Q. Wang, D. X. Shi, K. Ibrahim, J. Sun, Y. Wang, and H. J. Gao, "Epitaxial Growth of Flat Antimonene Monolayer: A New Honeycomb Analogue of Graphene," *Nano Lett.*, vol. 18, no. 3, pp. 2133–2139, 2018.

[85] G. Wang, R. Pandey, and S. P. Karna, "Atomically Thin Group V Elemental Films: Theoretical Investigations of Antimonene Allotropes," *ACS Appl. Mater. Interfaces*, vol. 7, no. 21, pp. 11490–11496, 2015.

[86] Y. Xu, B. Peng, H. Zhang, H. Shao, R. Zhang, and H. Zhu, "First-principle calculations of optical properties of monolayer arsenene and antimonene allotropes," *Ann. Phys.*, vol. 529, no. 4, p. 1600152, 2017.

[87] G. Wang, S. Higgins, K. Wang, D. Bennett, N. Milosavljevic, J. J. Magan, S. Zhang, X. Zhang, J. Wang, and W. J. Blau, "Intensity-dependent nonlinear refraction of antimonene dispersions in the visible and near-infrared region," *Appl. Opt.*, vol. 57, no. 22, pp. E147–E153, 2018.

[88] H. O. H. Churchill and P. Jarillo-herrero, "Two-Dimensional crystals:

Phosphorus joins the family," *Nat. Nanotechnol.*, vol. 9, no. 5, pp. 330–331, 2014.

[89] B. Guo, S. H. Wang, Z. X. Wu, Z. X. Wang, D. H. Wang, H. Huang, F. Zhang, Y. Q. Ge, and H. Zhang, "Sub-200 fs soliton mode-locked fiber laser based on bismuthene saturable absorber," *Opt. Express*, vol. 26, no. 18, pp. 22750–22760, 2018.

[90] X. Dai, Z. Qian, Q. Lin, L. Chen, R. Wang, and Y. Sun, "Benchmark investigation of band-gap tunability of monolayer semiconductors under hydrostatic pressure with focus-on antimony," *Nanomaterials*, vol. 10, p. 2154, 2020.

[91] Y. Huang, Y. H. Pan, R. Yang, L. H. Bao, L. Meng, H. L. Luo, Y. Q. Cai, G. D. Liu, W. J. Zhao, Z. Zhou, L. M. Wu, Z. L. Zhu, M. Huang, L. W. Liu, L. Liu, P. Cheng, K. H. Wu, S. B. Tian, C. Z. Gu, Y. G. Shi, Y. F. Guo, Z. G. Cheng, J. P. Hu, L. Zhao, G. H. Yang, E. Sutter, P. Sutter, Y. L. Wang, W. Ji, X. J. Zhou, and H. J. Gao, "Universal mechanical exfoliation of large-area 2D crystals," *Nat. Commun.*, vol. 11, p. 2453, 2020.

[92] X. Wu, Y. Shao, H. Liu, Z. Feng, Y. L. Wang, J. T. Sun, C. Liu, J. O. Wang, Z. L. Liu, S. Y. Zhu, Y. Q. Wang, S. X. Du, Y. G. Shi, K. Ibrahim, and H. J. Gao, "Epitaxial Growth and Air-Stability of Monolayer Antimonene on PdTe2," *Adv. Mater.*, vol. 29, no. 11, p. 1605407, 2017.

[93] M. Fortin-Deschênes, O. Waller, T. O. Menteş, A. Locatelli, S. Mukherjee, F. Genuzio, P. L. Levesque, A. Hébert, R. Martel, and O. Moutanabbir, "Synthesis of Antimonene on Germanium," *Nano Lett.*, vol. 17, no. 8, pp. 4970–4975, 2017.

[94] Y. Shao, Z. L. Liu, C. Cheng, X. Wu, H. Liu, C. Liu, J. Wang, S. Y. Zhu, Y. Q. Wang, D. X. Shi, K. Ibrahim, J. Sun, Y. Wang, and H. J. Gao, "Epitaxial growth of flat antimonene monolayer: a new honeycomb analogue of graphene," *Nano Lett.*, vol. 18, no. 3, pp. 2133–2139, 2018.

[95] X. Liu, I. Balla, H. Bergeron, G. P. Campbell, M. J. Bedzyk, and M. C. Hersam, "Rotationally commensurate growth of MoS$_2$ on epitaxial graphene," *ACS Nano*, vol. 10, pp. 1067–1075, 2016.

[96] J. Ji, X. Song, J. Liu, Z. Yan, C. Huo, S. Zhang, M. Su, L. Liao, W. Wang, Z. Ni, Y. Hao, and H. Zeng, "Two-dimensional antimonene single crystals grown by van der Waals epitaxy," *Nat. Commun.*, vol. 7, p. 13352, 2016.

[97] M. M. Ismail, J. Vigneshwaran, S. Arunbalaji, D. Mani, M. Arivanandhan, S. P. Jose, and R. Jayavel, "Antimonene nanosheets with enhanced electrochemical performance for energy storage applications," *Dalt. Trans.*, vol. 49, no. 39, pp. 13717–13725, 2020.

[98] H. Li, Q. Zhang, C. C. R. Yao, B. K. Tay, T. H. T. Edwin, A. Olivier, and D. Baillargeat, "From bulk to monolayer MoS$_2$: Evolution of Raman scattering," *Adv. Funct. Mater.*, vol. 22, no. 7, pp. 1385–1390, 2012.

[99] C. Lee, H. Yan, L. E. Brus, T. F. Heinz, J. Hone, and S. Ryu, "Anomalous

lattice vibrations of single- and few-layer MoS₂," *ACS Nano*, vol. 4, no. 5, pp. 2695–2700, 2010.

[100] L. Lu, X. Tang, R. Cao, L. Wu, Z. Li, G. Jing, B. Dong, S. Lu, Y. Li, Y. Xiang, J. Li, D. Fan, and H. Zhang, "Broadband Nonlinear Optical Response in Few-Layer Antimonene and Antimonene Quantum Dots: A Promising Optical Kerr Media with Enhanced Stability," *Adv. Opt. Mater.*, vol. 5, p. 1700301, 2017.

[101] G. Zhang, X. Tang, X. Fu, W. Chen, B. Shabbir, H. Zhang, Q. Liu, and M. Gong, "2D group-VA fluorinated antimonene: synthesis and saturable absorption," *Nanoscale*, vol. 11, p. 1762, 2019.

[102] G. Liu, F. Zhang, T. Wu, Z. Li, W. Zhang, K. Han, F. Xing, Z. Man, X. Ge, and S. Fu, "Single- and Dual-Wavelength Passively Mode-Locked Erbium-Doped Fiber Laser Based on Antimonene Saturable Absorber Based on Antimonene Saturable Absorber," *IEEE Photonics J.*, vol. 11, no. 3, p. 1503011, 2019.

[103] L. Li, J. G. Checkelsky, Y. S. Hor, C. Uher, A. F. Hebard, R. J. Cava, and N. P. Ong, "Phase Transitions of Dirac Electrons in Bismuth," *Science*, vol. 321, no. 5888, pp. 547–550, 2008.

[104] C. A. Hoffman, J. R. Meyer, F. J. Bartoli, A. D. Venere, X. J. Yi, C. L. Hou, H. C. Wang, J. B. Ketterson, and G. K. Wong, "Semimetal-to-semiconductor transition in bismuth thin films," *Phys. Rev. B*, vol. 48, no. 15, pp. 431–434, 1995.

[105] P. Hofmann, "The surfaces of bismuth: Structural and electronic properties," *Prog. Surf. Sci.*, vol. 81, pp. 191–245, 2006.

[106] T. Hirahara, I. Matsuda, S. Yamazaki, N. Miyata, S. Hasegawa, and T. Nagao, "Large surface-state conductivity in ultrathin Bi films Large surface-state conductivity in ultrathin Bi films," *Appl. Phys. Lett.*, vol. 91, p. 202106, 2007.

[107] A. Zhao and B. Wang, "Two-dimensional graphene-like Xenes as potential topological materials," *APL Mater.*, vol. 8, p. 030701, 2020.

[108] R. Bhuvaneswari, V. Nagarajan, and R. Chandiramouli, "Electronic properties of novel bismuthene nanosheets with adsorption studies of G-series nerve agent molecules – a DFT outlook," *Phys. Lett. A*, vol. 383, p. 125975, 2019.

[109] L. Lu, W. Wang, L. Wu, X. Jiang, Y. Xiang, J. Li, D. Fan, and H. Zhang, "All optical switching of two continuous waves in few layer bismuthene based on spatial cross-phase modulation," *ACS Photonics*, vol. 4, no. 11, pp. 2852–2861, 2017.

[110] L. Lu, Z. Liang, L. Wu, Y. Chen, Y. Song, S. C. Dhanabalan, J. S. Ponraj, B. Dong, Y Xiang, F. Xing, D. Fan, and H. Zhang, "Few-layer Bismuthene: Sonochemical Exfoliation, Nonlinear Optics and Applications for Ultrafast Photonics with Enhanced Stability," *Laser Photon. Rev.*, vol. 12, p. 1700221, 2018.

[111] Z. Liu, C. Liu, Y. Wu, W. Duan, F. Liu, and J. Wu, "Stable Nontrivial Z₂

Topology in Ultrathin Bi (111) Films: A First-Principles Study," *Phys. Rev. Lett.*, vol. 107, no. 13, p. 136805, 2011.

[112] C. Zhao, H. Zhang, X. Qi, Y. Chen, and Z. Wang, "Ultra-short pulse generation by a topological insulator based saturable absorber," *Appl. Phys. Lett.*, vol. 101, no. 21, p. 211106, 2012.

[113] C. Zhao, Y. Zou, Y. Chen, Z. Wang, S. Lu, H. Zhang, S. Wen, and D. Tang, "Wavelength-tunable picosecond soliton fiber laser with Topological Insulator: $Bi_2Se_3$ as a mode locker," *Opt. Express*, vol. 20, no. 25, pp. 27888–27895, 2012.

[114] S. Zhang, M. Xie, F. Li, Z. Yan, Y. Li, E. Kan, W. Liu, Z. Chen, and H. Zeng, "Semiconducting Group 15 Monolayers: A Broad Range of Band Gaps and High Carrier Mobilities Zuschriften," *Angew. Chemie*, vol. 128, no. 5, pp. 1698–1701, 2016.

[115] T. Feng, X. Li, T. Chai, P. Guo, Y. Zhang, R. Liu, J. Liu, J. Lu, and Y. Ge, "Bismuthene Nanosheets for 1 μm Multipulse Generation," *Langmuir*, vol. 36, pp. 3–8, 2020.

[116] T. Nagao, J. T. Sadowski, M. Saito, S. Yaginuma, Y. Fujikawa, T. Kogure, T. Ohno, Y. Hasegawa, S. Hasegawa, and T. Sakurai, "Nanofilm Allotrope and Phase Transformation of Ultrathin Bi Film on Si (111)-7 x 7," *Phys. Rev. Lett.*, vol. 93, no. 10, p. 105501, 2004.

[117] F. Reis, G. Li, L. Dudy, M. Bauernfeind, S. Glass, W. Hanke, R. Thomale, J. Schäfer, and R. Claessen, "Bismuthene on a SiC substrate: A candidate for a high-temperature quantum spin Hall material," *Science*, vol. 357, pp. 287–290, 2017.

[118] W. Gao, Z. Zheng, P. Wen, N. Huo, and J. Li, "Novel two-dimensional monoelemental and ternary materials: Growth, physics and application," *Nanophotonics*, vol. 9, no. 8, pp. 2147–2168, 2020.

[119] C. R. Ast and H. Höchst, "Fermi surface of bi(111) measured by photoemission spectroscopy," *Phys. Rev. Lett.*, vol. 87, no. 17, pp. 20–23, 2001.

[120] V. Chis, G. Benedek, P. M. Echenique, and E. V. Chulkov, "Phonons in ultrathin Bi(111) films: Role of spin-orbit coupling in electron-phonon interaction," *Phys. Rev. B*, vol. 87, no. 7, p. 075412, 2013.

[121] P. Guo, X. Li, T. Chai, T. Feng, and Y. Ge, "Few-layer bismuthene for robust ultrafast photonics in C-Band optical communications," *Nat. Nanotechnol.*, vol. 30, p. 354002, 2019.

[122] P. Guo, X. Li, T. Feng, Y. Zhang, and W. Xu, "Few-layer bismuthene for coexistence of harmonic and dual-wavelength in a mode-locked fiber laser," *ACS Appl. Mater. Interfaces*, vol. 12, no. 28, pp. 31757–31763, 2020.

[123] F. Yang, A. O. Elnabawy, R. Schimmenti, P. Song, J. Wang, Z. Peng, S. Yao, R. Deng, S. Song, Y. Lin, M. Mavrikakis, and W. Xu, "Bismuthene for highly efficient carbon dioxide electroreduction reaction," *Nat. Commun.*, vol. 11, p.

1088, 2020.

[124] T. Chai, X. Li, T. Feng, P. Guo, Y. Song, Y. Chen, and H. Zhang, "Few-layer bismuthene for ultrashort pulse generation in dissipative system based on evanescent field," *Nanoscale*, vol. 10, pp. 17617–17622, 2018.

[125] T. Feng, X. Li, T. Chai, P. Guo, Y. Zhang, R. Liu, J. Liu, J. Lu, and Y. Ge, "Few-layer bismuthene for 1-μm ultrafast laser applications," *Beilsten Arch.*, vol. 1, p. 28, 2019.

[126] K. Y. Lau and D. Hou, "Recent research and advances of material-based saturable absorber in mode-locked fiber laser," *Opt. Laser Technol.*, vol. 137, p. 106826, 2021.

[127] N. H. Park, H. Jeong, S. Y. Choi, M. H. Kim, and F. Rotermund, "Monolayer graphene saturable absorbers with strongly enhanced evanescent-field interaction for ultrafast fiber laser mode-locking," *Opt. Express*, vol. 23, no. 15, pp. 19806–19812, 2015.

[128] L. Li, Y. Su, Y. Wang, X. Wang, Y. Wang, X. Li, D. Mao, and J. Si, "Femtosecond Passively Er-Doped Mode-Locked Fiber Laser With $WS_2$ Solution Saturable Absorber," *IEEE J. Sel. Top. Quantum Electron.*, vol. 23, no. 1, p. 1100306, 2017.

[129] L. Huang, Y. Zhang, and X. Liu, "Dynamics of carbon nanotube-based mode-locking fiber lasers," *Nanophotonics*, vol. 9, no. 9, pp. 2731–2761, 2020.

[130] R. Khazaeinezhad, S. H. Kassani, H. Jeong, T. Nazari, D.-I. Yeom, and K. Oh, "Mode-Locked All-Fiber Lasers at Both Anomalous and Normal Dispersion Regimes Based on Spin-Coated $MoS_2$ Nano-Sheets on a Side-Polished Fiber," *IEEE Photonics J.*, vol. 7, no. 1, p. 1500109, 2015.

[131] E. K. Ng, K. Y. Lau, H. K. Lee, N. M. Yusoff, A. R. Sarmani, M. F. Omar, and M. A. Mahdi, "L-band femtosecond fiber laser based on a reduced graphene oxide polymer composite saturable absorber," *Opt. Mater. Express*, vol. 11, no. 1, pp. 59–72, 2021.

[132] E. K. Ng, K. Y. Lau, H. K. Lee, M. H. Abu Bakar, Y. M. Kamil, M. F. Omar, and M. A. Mahdi, "Saturable absorber incorporating graphene oxide polymer composite through dip coating for mode-locked fiber laser," *Opt. Mater.*, vol. 100, p. 109619, 2020.

[133] J. Du, Q. Wang, G. Jiang, C. Xu, C. Zhao, Y. Xiang, Y. Chen, S. Wen, and H. Zhang, "Ytterbium-doped fiber laser passively mode locked by few-layer molybdenum disulfide ($MoS_2$) saturable absorber functioned with evanescent field interaction," *Sci. Rep.*, vol. 4, p. 6346, 2014.

[134] J. Wang, H. Chen, Z. Jiang, J. Yin, J. Wang, M. ZHang, T. He, J. Li, P. Yan, and S. Ruan, "Mode-locked thulium-doped fiber laser with chemical vapor deposited molybdenum ditelluride," *Opt. Lett.*, vol. 43, no. 9, pp. 1998–2001, 2018.

[135] K. Kieu and M. Mansuripur, "Femtosecond laser pulse generation with a fiber

[136] C. Wang, L. Wang, X. Li, W. Luo, T. Feng, Y. Zhang, P. Guo, and Y. Ge, "Few-layer bismuthene for femtosecond soliton molecules generation in Er-doped fiber laser," *Nanotechnology*, vol. 30, p. 025504, 2019.

[137] K. Y. Lau, M. H. Abu Bakar, F. D. Muhammad, A. A. Latif, M. F. Omar, Z. Yusoff, and M. A. Mahdi, "Dual-wavelength, mode-locked erbium-doped fiber laser employing a graphene/polymethyl-methacrylate saturable absorber," *Opt. Express*, vol. 26, no. 10, pp. 12790–12800, 2018.

[138] K. Y. Lau, N. H. Zainol Abidin, M. H. Abu Bakar, A. A. Latif, F. D. Muhammad, N. M. Huang, M. F. Omar, and M. A. Mahdi, "Passively mode-locked ultrashort pulse fiber laser incorporating multi-layered graphene nanoplatelets saturable absorber," *J. Phys. Commun.*, vol. 2, p. 075005, 2018.

[139] Q. Bao, H. Zhang, Z. Ni, Y. Wang, L. Polavarapu, Z. Shen, Q. H. Xu, D. Tang, and K. P. Loh, "Monolayer graphene as a saturable absorber in a mode-locked laser," *Nano Res.*, vol. 4, no. 3, pp. 297–307, 2011.

[140] C. Hoon Kwak, L. Yeung Lak, and K. Seong Gyu, "Analysis of asymmetric Z-scan measurement for large optical nonlinearities in an amorphous $As_2S_3$ thin film," *J. Opt. Soc. Am. B*, vol. 16, no. 4, pp. 600–604, 1999.

[141] G.-R. Lin, I.-H. Chiu, and M.-C. Wu, "1.2-ps mode-locked semiconductor optical amplifier fiber laser pulses generated by 60-ps backward dark-optical comb injection and soliton compression," *Opt. Express*, vol. 13, no. 3, pp. 1008–1014, 2005.

[142] P. Lazaridis, G. Debarge, and P. Gallion, "Time-bandwidth product of chirped $sech^2$ pulses: application to phase-amplitude-coupling factor measurement," *Opt. Lett.*, vol. 20, no. 10, pp. 1160–1162, 1995.

[143] A. Carvalho and A. H. C. Neto, "Phosphorene: Overcoming the Oxidation Barrier," *ACS Cent. Sci.*, vol. 1, pp. 289–291, 2015.

[144] R. A. Doganov, E. C. T. O'Farrell, S. P. Koenig, Y. Yeo, A. Ziletti, A. Carvalho, D. K. Campbell, D. F. Coker, K. Watanabe, T. Taniguchi, A. H. Castro Neto, and B. Özyilmaz, "Transport properties of pristine few-layer black phosphorus by van der Waals passivation in an inert atmosphere," *Nat. Commun.*, vol. 6, p. 6647, 2015.

[145] W. Zhu, M. N. Yogeesh, S. Yang, S. H. Aldave, J. Kim, S. S. Sonde, L. Tao, N. Lu, and D. Akinwande, "Flexible black phosphorus ambipolar transistors, circuits and AM demodulator," *Nano Lett.*, vol. 15, no. 3, pp. 1883–1890, 2015.

[146] A. Avsar, I. J. Vera-Marun, J. Y. Tan, K. Watanabe, T. Taniguchi, A. H. C. Neto, and B. Özyilmaz, "Air-Stable Transport in Graphene-Contaced, Fully Encapsulated Ultrathin Black Phosphorous-Based Field-Effect Transistors," *ACS Nano*, vol. 9, no. 4, pp. 4138–4145, 2015.

[147] O. Traxer and E. X. Keller, "Thulium fiber laser: the new player for kidney stone treatment? A comparison with Holmium:YAG laser," *World J. Urol.*, vol. 38, no. 8, pp. 1883–1894, 2020.

[148] M. Amin, M. Farhat, and H. Bağcı, "An ultra-broadband multilayered graphene absorber," *Opt. Express*, vol. 21, no. 24, pp. 29938–29948, 2013.

[149] X. Li, G. Feng, and S. Lin, "Ultra-wideband terahertz absorber based on graphene modulation," *Appl. Opt.*, vol. 60, no. 11, pp. 3170–3175, 2021.

[150] Y. Cui and X. Liu, "Revelation of the birth and extinction dynamics of solitons in SWNT-mode-locked fiber lasers," *Photonics Res.*, vol. 7, no. 4, pp. 423–430, 2019.

[151] X. Liu, X. Yao, and Y. Cui, "Real-Time Observation of the Buildup of Soliton Molecules," *Phys. Rev. Lett.*, vol. 121, p. 023905, 2018.

[152] Y. Zhang, L. Huang, Y. Cui, and X. Liu, "Unveiling external motion dynamics of solitons in passively mode-locked fiber lasers," *Opt. Lett.*, vol. 45, no. 17, pp. 4835-4838, 2020.

[153] Y. Zhang, Y. Cui, L. Huang, L. Tong, and X. Liu, "Full-field real-time characterization of creeping solitons dynamics in a mode-locked fiber laser," *Opt. Lett.*, vol. 45, no. 22, pp. 6246–6249, 2020.

[154] P. Yan, R. Lin, S. Ruan, A. Liu, H. Chen, Y. Zheng, S. Chen, C. Guo, and H. Hu, "A practical topological insulator saturable absorber for mode-locked fiber laser," *Sci. Rep.*, vol. 5, p. 8690, 2015.